\documentclass[10pt,journal,compsoc]{IEEEtran}

\ifCLASSOPTIONcompsoc
\usepackage[nocompress]{cite}
\else
\usepackage{cite}
\fi

\usepackage{soul}
\usepackage{listings}
\usepackage{balance}
\usepackage{algorithm}
\usepackage[noend]{algpseudocode}
\algnewcommand\And{\textbf{ and }}
\algnewcommand\Not{\textbf{not }}
\usepackage{graphicx}
\usepackage[caption=false]{subfig}
\usepackage{color}
\usepackage{multirow}
\usepackage{textcomp}
\usepackage{varwidth}
\usepackage{stackengine,graphicx,xcolor}
\usepackage{tikz}
\usepackage{booktabs}
\usepackage{paralist}
\usepackage[T1]{fontenc}
\usepackage{array}
\usepackage{xcolor,colortbl}
\usepackage{fancyvrb}
\usepackage{url}
\usepackage{wrapfig}
\usepackage{adjustbox}

\newcommand{\ra}[1]{\renewcommand{\arraystretch}{#1}}

\newcolumntype{L}[1]{>{\raggedright\let\newline\\\arraybackslash\hspace{0pt}}m{#1}}
\newcolumntype{C}[1]{>{\centering\let\newline\\\arraybackslash\hspace{0pt}}m{#1}}
\newcolumntype{R}[1]{>{\raggedleft\let\newline\\\arraybackslash\hspace{0pt}}m{#1}}

\definecolor{listinggray}{gray}{0.9}
\definecolor{lbcolor}{rgb}{0.9,0.9,0.9}

\usepackage{soul}
\newcommand{\review}[1]{\textcolor{black}{#1}}
\newcommand{\reviewR}[1]{\textcolor{black}{#1}}
\newcommand{\optional}[1]{}

\begin{document}
\setlength\belowcaptionskip{-10pt}
\setlength\abovecaptionskip{0pt}

\title{DMRlib: Easy-coding and Efficient Resource Management for Job Malleability}

\author{
Sergio~Iserte, Rafael~Mayo, Enrique~S.~Quintana-Ort\'i and Antonio~J.~Pe\~na

\IEEEcompsocitemizethanks
{

\IEEEcompsocthanksitem S. Iserte is with the Dept. of Mechanical and Engineering Construction at Universitat Jaume I (UJI), Spain.\hfil
\IEEEcompsocthanksitem R. Mayo is with the Dept. of Computer Science and Engineering at UJI.\hfil
\IEEEcompsocthanksitem E. S. Quintana-Ort\'i is with Dept. of Computing Engineering at Universitat Polit\`ecnica de Val\`encia (UPV), Spain.\hfil
\IEEEcompsocthanksitem A. J. Pe\~na is with the Barcelona Supercomputing Center (BSC).}
}


\IEEEtitleabstractindextext{%
\begin{abstract}
Process malleability has proved to have a highly positive impact on the resource utilization and global productivity in data centers compared with the conventional static resource allocation policy. However, the non-negligible additional development effort this solution imposes has constrained its adoption by the scientific programming community.
In this work, we present DMRlib, a library designed to offer the global advantages of process malleability while providing a minimalist MPI-like syntax.
The library includes a series of predefined communication patterns that greatly ease the development of malleable applications.
In addition, we deploy several scenarios to demonstrate the positive impact of process malleability featuring different scalability patterns.
Concretely, we study two job submission modes (rigid and moldable) in order to identify the best-case scenarios for malleability using metrics such as resource allocation rate, completed jobs per second, and energy consumption.
The experiments prove that our elastic approach may improve global throughput by a factor higher than 3x compared to the traditional workloads of non-malleable jobs.
\end{abstract}

}

\maketitle

\IEEEdisplaynontitleabstractindextext

\IEEEraisesectionheading{\section{Introduction}\label{sec:introduction}}
Traditionally, parallel applications running in high-performance computing (HPC) facilities allocate all resources during their entire execution period. That would not be a remarkable issue if scaling up a job reduced its completion time, enabling those resources to become available earlier for other jobs.
In practice, however, the purpose of scaling up a job is very likely to compute over a larger problem, which will need a longer time to terminate~\cite{Lublin2003} and, consequently, take longer to release its resources.
Hence, far from solving the problem earlier (strong scaling), the job will need additional resources during a longer time in order to solve the new problem size (weak scaling).
In this scenario, a few jobs may monopolize a large number of resources most of the time, preventing other jobs from being initiated.

\reviewR{
In this work, malleability refers to the specific case of elasticity in the number of workers. In this regard, malleable workloads take care of checking the status of the HPC facility in terms of available resources.  Malleable workloads are defined as a combination of the following four actors working together:}
\begin{inparaenum}[i)]
 \item user applications with support for on-the-fly scale-up/down (process malleability);
 \item a parallel distributed runtime (PDR) responsible for re-scaling the jobs;
 \item a resource management system (RMS) with the necessary logic to reallocate resources considering the cluster status; and
 \item a communication mechanism that allows i)-iii) to interact in order to perform the job reconfiguration actions.
\end{inparaenum}
\newline
\indent
\reviewR{The main target of process malleability is iterative applications since they present clear processes synchronization points where resizes can be easily triggered. Nevertheless, we can find resizing in different types
of applications, such as master/worker schemes in~\cite{Compres2016} and~\cite{Iserte2018hpg}.}
Malleable jobs can re-scale themselves at execution time by expansion or shrinkage. Moreover, a system-aware job scheduler handles workload information, including, among others, resource utilization or job status.
Hence, readjusting the workload taking into account the cluster status benefits not only the HPC facility---which may experience an increase in the number of completed jobs per second (global throughput) and a higher utilization of resources---but also the end-users, who may enjoy shorter response (waiting plus execution) times for their jobs.
\newline
\indent
However, despite the advantages of job malleability and a variety of existing tools for this purpose (see Section~\ref{sec:related}), application developers are still reluctant to integrate malleability in their codes, mainly because those solutions require a considerable re-coding effort which may even include changing the programming paradigm.
\newline
\indent
This paper presents the dynamic management of resources library (DMRlib), an effort to promote and publicize the benefits of malleability in production environments (see Section~\ref{sec:dmrlib}). DMRlib exposes a minimalist set of semantics in an MPI-like  syntax (see Appendix~\ref{app:api}), which ease malleability adoption.
Specifically, DMRlib enables those developers familiar with the MPI programming model to easily integrate malleability mechanisms in their applications by using the standard MPI communication routines and a reconfiguration trigger, which will determine the synchronization point for malleability actions.
\newline
\indent
DMRlib improves the state-of-the-art in programming models and runtimes as detailed in Section~\ref{sec:usability}, which presents a thorough study of the available malleability solutions. The analysis there compares the semantics offered by different solutions under fair conditions. For this purpose, the study is opened with the implementation of a generic migration of processes. Then, the characteristics of the frameworks are evaluated and compared, highlighting their strengths and flaws. The section ends showcasing the practical experience of how to adopt malleability with DMRlib in a set of applications, further evaluated in Section~\ref{sec:perf_eval}.
\newline\indent
In this work, a job reconfiguration is encapsulated in a single call to DMRlib that abstracts the resource reallocation in Slurm, the process management (spawns and terminations), and the data redistribution among processes.
In those cases where our library does not support the data redistribution, users can provide their own redistribution functions using any function implemented in the underlying MPI library.
We consider our solution as ``MPI-friendly'' since it does not include new functions or wrappers to the MPI standard, unlike other approaches.

Using DMRlib, we have developed four malleable applications, which are then leveraged to assess the behavior of moldable and malleable jobs within a workload.
\reviewR{Section~\ref{sec:perf_eval} proves that}, at the cost of a small extra coding effort, it is possible to obtain malleable workloads that yield:
\begin{inparaenum}[i)]
    \item significantly higher system throughput than their traditional counterpart;
    \item reduction in the completion time of each particular application; and
    \item lower energy consumption to process the workload (see Appendix~\ref{app:energy}).
\end{inparaenum}
We conclude this analysis with a study on the impact of malleability when not all the jobs in the workload can be resized.

\reviewR{In summary, the main contributions of this paper are:
\begin{itemize}
\item A minimalist set of semantics following an MPI-like syntax, which eases malleability.
\item A usability-based study of the available malleability solutions, highlighting their strengths and weaknesses via a simple showcase. 
\item A practical proof that DMRlib enables obtaining malleable workloads that increase system productivity.
\end{itemize}}
\reviewR{Finally, the paper ends in Section~\ref{sec:conclusions} with the most relevant conclusions extracted from this work.}

\section{Related Work}
\label{sec:related}
In this section, we briefly review several related approaches, categorized as on-disk and in-memory data redistributions, and system-aware reconfigurations.

\subsection{On-Disk Reconfiguration}
On-disk reconfiguration stores/loads the state of a job \mbox{in/from} non-volatile memory.
The checkpoint/restart (C/R) mechanism is the most famous instance of this strategy.
C/R stores the state of an application at a given point of its execution and recovers it at a later time.
Traditionally, it is used for preventing data loss in case of a system fault.
However, C/R has also been leveraged in job malleability as a means of halting the execution and resuming it with a different number of processes.
\review{
Examples: 
the malleability extensions of the MPI libraries Process Checkpointing and Migration (PCM)~\cite{ElMaghraoui2007}, and Scalable Checkpoint\slash Restart (SCR)~\cite{Lemarinier2016}; CHARM++ with its migratable objects called \textit{chares}~\cite{Acun2014}, and 
Adaptive MPI (AMPI)~\cite{Gupta} that implements MPI on top of CHARM++'s adaptive runtime system.
}

\subsection{In-Memory Reconfiguration}
Traditional on-disk C/R solutions attain low performance because of the cost of disk.
In-memory C/R solutions palliate this undesirable overhead~\cite{Zheng2012}, transferring the data among processes, point--to--point or collectively, without accessing the disk.
The overhead of the reconfiguration provided by traditional C/R mechanisms compared with a dynamic redistribution of data is discussed in~\cite{Lemarinier2016},~\cite{Iserte-thesis}.
\review{
Examples of in-memory reconfiguration are: EasyGrid Application Management System (AMS)~\cite{Ribeiro2013}; Flex-MPI~\cite{Martin2013}; and the User Level Failure Migration (ULFM) MPI malleability extension~\cite{Lemarinier2016}.
}

\subsection{System-Aware Reconfiguration}
The previous two subsections list a collection of libraries and runtimes capable of reconfiguring applications using different approaches.
Those solutions are not focused on system-aware reconfiguration and do not make an effort to work side by side with the RMS.
Instead, the previous solutions implement a simple ad-hoc scheduler, which triggers the reconfiguration actions when necessary. 
In this section, we present the most relevant efforts in job malleability aimed at adaptive workloads, ready to be adopted in production environments. 
These integrate reconfiguration capabilities with an RMS which is aware of the cluster status.

ReSHAPE~\cite{Sudarsan2007} is a coupled solution for adaptive workloads that includes its specific reconfiguration libraries, scheduler, and runtime system.
This strong integration forces ReSHAPE users to develop applications that are compatible with this system.

The power-aware resource manager (PARM)~\cite{Sarood2014} uses over-provisioning, power capping, and job malleability to maximize job throughput under a strict power budget in over-provisioned facilities.
Regarding the malleability, it relies on the CHARM++ runtime support, which dynamically redistributes compute objects to processors.

The work developed in~\cite{Prabhakaran2015} combines AMPI with the job scheduler Torque/Maui to tackle malleable jobs. In this regard, this solution establishes a communication layer between the CHARM++ runtime and Torque/Maui.

\reviewR{Elastic MPI~\cite{Compres2016} is an infrastructure and a set of API extensions for malleable execution of MPI applications based on Slurm and MPICH.}
Using the functions provided by this API, an application is declared as malleable and, periodically, the processes of the application \reviewR{checks} whether Slurm has initiated a reconfiguration.
MPICH has been enhanced with a new set of functions that replace and complement the standard implementation.
In addition to being MPICH-dependent (support for other MPI implementations is not presented), this approach does not assist in data redistribution.

The Dynamic Management of Resources (DMR) API~\cite{Iserte2017} implements a communication layer between the OmpSs runtime (Nanos++) and Slurm that allows MPI applications to be resized.
The DMR~API relies on the off-load semantics of OmpSs~\cite{Sainz2015} for automatically handling processes and data redistribution, and on Slurm for managing and reallocating resources.
Although the DMR~API provides a highly usable interface, irregular applications (e.g., applications implementing a consumer--producer scheme) require a special effort to integrate the reconfiguration capabilities because not every process features the same data structures.
An additional disadvantage appears for object-oriented applications, such as those leveraging C++ classes, whose code has to be refactored.
Concretely, the data utilized in the functions cannot be conveyed to them as class members, but have to be passed as parameters of the functions.

Thanks to these efforts, malleability is becoming an easily accessible technology.
As it could be expected, the solutions incur overhead when redistributing data among processes by loading data from disk or transferring it through the network interconnect.
However, regarding the scheduling overhead of solutions that establish communication with an RMS, it has been reported as virtually null.

\review{
Despite the wide variety of reconfiguration frameworks, none of them combines all the features that should have a ubiquitous solution:
\begin{itemize}
    \item Avoid disk reads/writes for redistributing user data among processes.
    \item Offer an API familiar to the vast majority of HPC developers, such as MPI.
    \item Be compatible with any MPI standard implementation.
    \item Be integrated with well-accepted RMS in HPC.
\end{itemize}
In order to address this situation, in this paper we present DMRlib, a novel framework that addresses malleabilty adoption from a different perspective.
In detail, Section~\ref{sec:usability} studies and evaluates these frameworks, demonstrating that, before DMRlib, the most practical approaches were based on the semantics of CHARM++ or OmpSs. In those approaches, the code-writing process requires less effort; however, prior to adopting any of these solutions, the users are expected to learn the specifics of each programming model and their particular syntax.
In summary, the handicaps of adopting malleability remain in that the described frameworks are firmly bound to specific programming models, or require considerable efforts from application developers.
DMRlib aims to overcome those limitations providing a minimal set of powerful tools, which facilitate the adoption of a useful and interesting mechanism such is malleability. 
}

\section{DMR\lowercase{lib}}
\label{sec:dmrlib}
\begin{figure}
  \centering
  \includegraphics[clip,width=0.85\columnwidth] {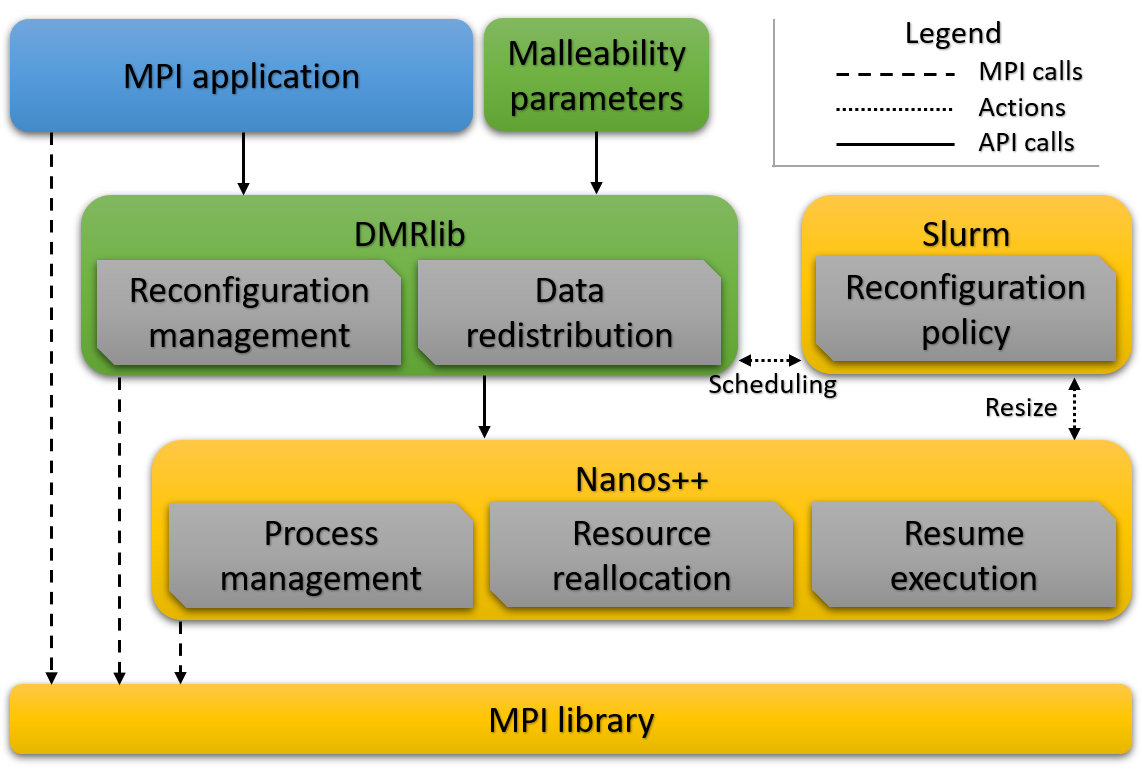}
  \caption{Execution environment of a malleable application using DMRlib.}\label{fig:dmrlib}
\end{figure}

\begin{figure}
    \centering
    \includegraphics[clip,width=0.8\columnwidth,trim={0cm 8.2cm 0cm 0cm}] {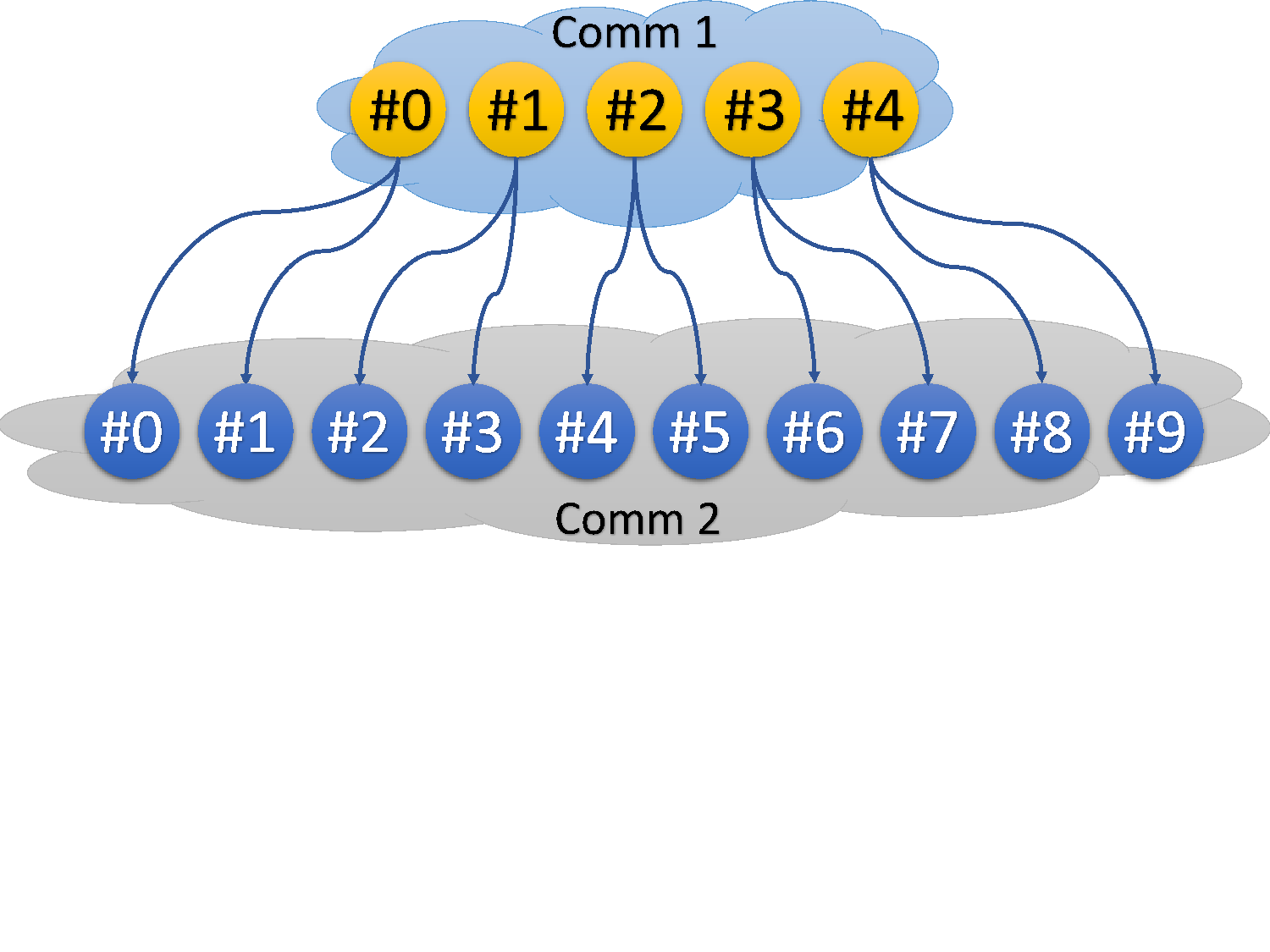}
    \caption{Example of an expansion from 5 to 10 processes.}\label{fig:expand}
\end{figure}

Our library for dynamic management of resources (DMRlib) is designed to facilitate the adoption of malleability to application developers.
Based on the DMR~API~\cite{Iserte2018}, DMRlib provides a higher-level abstraction layer that orchestrates the interactions between application, runtime, and RMS.
Figure~\ref{fig:dmrlib} depicts those interactions and shows how are communicated the different modules with the original MPI user application. Thus,
\review{
DMRlib initiates the reconfiguration, honoring the malleability parameters provided by the user, by communicating with Slurm.
\reviewR{
Then, DMRlib creates the new processes, using the standard MPI function \texttt{MPI\_Comm\_spawn}, and redistributes the data among them, using for example point-to-point MPI functions such as MPI\_Send or MPI\ Recv; collectives such as MPI\_Scatter or MPI\_Gather; or the predefined redistribution functions provided by DMRlib (see Section~\ref{subsec:patterns}).}
Finally, Nanos++ (the OmpSs runtime) takes control of the reconfiguration, handling the termination of the initial processes, reallocating resources, and resuming the execution on the new processes, at the same point where the malleability action was triggered. 
This feature is regarded as one of the most distinctive features of the family of DMR solutions.
\newline\indent
Later in the paper, in Section~\ref{subsec:sched}, we describe the utilized malleability criteria, which determines the job reconfiguration action. 
With this policy, aimed to improve the global system productivity, jobs are expanded or shrunk to improve resource utilization or reduce the pending jobs' waiting time.
}
Figure~\ref{fig:expand} illustrates an example with an expansion from 5 to 10 processes.
Henceforth we call \textbf{parents} to the processes in the initial state of a malleable operation and \textbf{child} to each of the final spawned processes.
The figure shows five parent processes that are expanded to 10 child processes. 
When a malleable operation is scheduled, it creates a new communicator (``Comm 2'' in the figure) that will be used by the child processes.
Once the latter has been created, communication occurs between parent and child processes through the corresponding intercommunicator.
This communication poses a robust restart where the user data, as well as the reconfiguration information, is sent from parents to children (as the arrows represent in the figure.)

DMRlib performs communication with Nanos++ and the RMS.
This work adopts a version of Slurm with support for malleability, also used by the DMR~API.

\subsection{Main Procedure}
DMRlib's main procedure triggers and subsequently handles the complete reconfiguration process. Listing~\ref{code:macro-header} shows the usage of this procedure, which receives five arguments corresponding to function names:
\begin{itemize}
\item[\texttt{compute}:] the function that will be executed when the reconfiguration procedure ends and the child processes resume the execution of the application. 
Typically, in an iterative application, this function is the same that invokes the reconfiguration.
\item[\texttt{send\_expand}:] the function executed by the parent processes in \textit{Comm~1} when performing an expansion. 
This function implements the algorithm for sending data from the parent processes to the child processes.
\item[\texttt{recv\_expand}:] the function executed by the child processes in \textit{Comm~2} when performing an expansion. 
This function implements the algorithm for receiving data in the child processes sent from the parent processes.
\item[\texttt{send\_shrink}:] similar to \texttt{send\_expand} but for shrinking.
\item[\texttt{recv\_shrink}:] similar to \texttt{recv\_expand} but for shrinking.
\end{itemize}

\begin{lstlisting}[float,caption=Reconfiguration macro definition in DMRlib., label=code:macro-header, captionpos=b, numbers=none]
#define DMR_RECONFIG(compute, send_expand, recv_expand, send_shrink, recv_shrink)
\end{lstlisting}

Algorithm~\ref{alg:macro-body} details the reconfiguration procedure.
First, the current stage of the reconfiguration is checked (line~1) to determine the role (parents or children) of the invoking processes.
For this purpose, the library tries to retrieve the parent communicator.
If there is one (line~2), it will be used to handle the data redistribution.
Line~3 determines whether the resize action is an expansion or a shrinkage by comparing the number of processes in both communicators. Thus, if the current global communicator (\texttt{MPI\_COMM\_WORLD}) contains a higher number of processes than the parent communicator, the library invokes the user function for receiving data in an expansion (line~4);
otherwise, it calls the receiving data function for a shrinkage (line~6).

\begin{algorithm}[t]
    \scriptsize
\caption{Reconfiguration Procedure}\label{alg:macro-body}
\begin{algorithmic}[1]
\State $parentComm \gets \textit{MPI\_Comm\_get\_parent()}$
\If {$parentComm \neq \texttt{MPI\_COMM\_NULL}$}
\If {$commWorldSize > parentCommSize$}
\State $\textit{recv\_expand()}$
\Else
\State $\textit{recv\_shrink()}$
\EndIf
\State $\textit{MPI\_Comm\_disconnect(}parentComm\textit{)}$
\Else
\State $action \gets \textit{DMR\_Reconfiguration(}\&newComm\textit{)}$
\If {$action = \texttt{expand}$}
\State $factor \gets newComm\_size / comm\_size$
      \For{$i\gets 1, factor$}
        \State $dstRank \gets myRank \times factor + i$

     \State \begin{varwidth}[t]{\columnwidth}
    $\texttt{\#pragma omp task onto(}newComm\texttt{, }dstRank\texttt{)}$
      \end{varwidth}

        \State \textit{compute()}
      \EndFor
\State $\textit{send\_expand()}$
\State $\textit{DMR\_Detach()}$
\Else
\If {$action = \texttt{shrink}$}
\State $factor \gets commSize / newCommSize$
\State $dstRank \gets myRank / factor$

    \State \begin{varwidth}[t]{\columnwidth}
    $\texttt{\#pragma omp task onto(}newComm\texttt{, }dstRank\texttt{)}$
      \end{varwidth}

\State $\textit{compute()}$
\State $\textit{send\_shrink()}$
\State $\textit{DMR\_Detach()}$
\Else
\State \texttt{pass}
\Comment{No action has been scheduled.}
\EndIf
\EndIf
\EndIf
\end{algorithmic}
\end{algorithm}

\reviewR{
In line~1, no reconfiguration is ongoing if \texttt{MPI\_Comm\_get\_parent} returns \textit{MPI\_COMM\_NULL}, since the current communicator has no parent. In other words, a new resize may be performed. For this reason, 
the application, in line~9, communicates its readiness for being resized to the runtime.}
At this point, \textit{action} may receive three values: ``expand'', ``shrink'' or ``none''.
In the latter case, the program execution continues normally (line~27); otherwise, the macro performs the job re-scaling.

\reviewR{In the case of an expansion the scalability factor is calculated (line 11)  to determine the number of links to be established by each initial process.}
With this \textit{factor}, each process calculates the ranks of its peer processes in the new communicator (line 13) and establishes the communication to identify the function where the execution will be resumed after the reconfiguration (lines 14 and 15). In line~16, the data is sent utilizing the function provided by the user: \texttt{send\_expand}.
Finally, the new processes are disconnected from \textit{Comm~1} and can continue the execution of the application in a new computational step.
Moreover, the initial processes terminate their execution (line~17).

In the case of a shrinkage (line~19), the procedure is similar to an expansion.
DMRlib calculates the communication \textit{factor} (line~20), establishes the communication channels among the processes in the different communicators, and sets the resuming point for the execution (line~23).
Finally, the data is sent using the function \texttt{send\_shrink}, and the processes are detached from the initial communicator (line~25).

\subsection{Parametrization}
DMRlib includes routines (see Appendix~\ref{app:api}) to customize job malleability fully.
For instance, a user can utilize the function \texttt{DMR\_Set\_parameters} to define the boundaries of malleability in terms of the number of processes.
The arguments passed to this function are:
\begin{itemize}
\item \texttt{min}: minimum number of processes to run the job---the lower limit for malleability.
\item \texttt{max}: maximum number of processes to run the job---the upper limit for malleability.
\item \texttt{pref}: preferred number of processes to run the job.
Although optional, it allows the reconfiguration policy in Slurm to take more convenient actions.
\end{itemize}

\review{
The overhead of malleability frameworks usually has two different sources: reconfiguration scheduling and data transfer. In previous work, we already demonstrated that the overhead introduced by the DMR~API was low~\cite{Iserte2018}. Our conclusions indicated that:
\begin{itemize}
    \item Overhead is \reviewR{dominated by the data size} to transfer.
    \item Malleability scheduling time is negligible.
    \item MPI process spawn and destroy operation time depends on the number of processes and the MPI library.
    \item Interconnection network \reviewR{bandwidth} is crucial.
\end{itemize}
}

Moreover, in applications that perform short-step executions---that is, computational steps that only take a few milliseconds---triggering a reconfiguration at every step may generate a substantial overhead, since the time taken by the computational step may be in the same order of magnitude as the time needed for the reconfiguration scheduling.
For this reason, DMRlib implements two mechanisms for ignoring scheduling reconfigurations: during a given period (\texttt{DMR\_Set\_sched\_period}) and for a determined number of steps (\texttt{DMR\_Set\_sched\_iteration}).

\subsection{Usage}
In order to turn an application into malleable, a user has to include the macro \texttt{DMR\_RECONFIG} in the code, as illustrated in Listing~\ref{code:dmrlib-usage}.
This excerpt of code shows the function containing the main loop (line~1) and hence specifies the point where malleability will occur.
This function has three parameters: the data structure pointer, its size, and the current iteration.
Initially, the malleability limits are configured (line~2). They can be modified at any time to meet the requirements of different computational stages.

At the beginning of each iteration of the main loop (line~3), \texttt{DMR\_RECONFIG} checks whether a reconfiguration is ongoing or if the RMS can improve the system status by resizing the job via DMRlib (line~4).
The macro must specify the appropriate functions for the reconfiguration: the first is the invoking function itself, while the remaining four arguments are user-defined redistribution function calls.
The ``receiving'' functions in the macro (\textit{recv\_expand} and \textit{recv\_shrink}) operate with the memory address instead of the pointer (or the value itself for the \textit{data\_size} variable).
The reason is that these functions are invoked just after the new processes have been spawned, and they have to allocate new memory to accommodate the data.
The rest of the functions remain unaltered (line~5).

\begin{lstlisting}[float,caption=Enabling malleability using DMRlib in a user code., label=code:dmrlib-usage, captionpos=b]
void compute(double *data, int data_size, int step) {
    DMR_Set_parameters(min, max, pref);
    for (int i = step; i < TOTAL_STEPS; i++) {
        DMR_RECONFIG(compute(data, data_size, i), send_expand(data, data_size), recv_expand(&data, &data_size), send_shrink(data, data_size), recv_shrink(&data, &data_size));
        /* Computation */
    }
}
\end{lstlisting}

Listing~\ref{code:dmrlib-user-expansion} shows an example of the redistribution functions used in the previous macro for an expansion action (\texttt{send\_ex\-pand} and \texttt{recv\_ex\-pand}).
Here we describe a case similar to that shown in Figure~\ref{fig:expand}; that is, data transfers due to an expansion to an integer multiple of the initial number of processes.

The first function (line~1) will be executed by the parent processes, which are in charge of sending the data.
Dividing the number of child processes by the number of parent processes returns the scalability factor of this expansion (line~2).
With this value, we calculate the size of the data chunks (line~3).
Furthermore, \textit{factor} is used to determine the number of \textit{send} operations to be performed by each original process (line~4) and the rank in the new communicator that identifies the destination of each chunk of data (line~5).
In line~6, we use the standard \texttt{MPI\_Send} function, passing the buffer pointer to the appropriate data chunk (argument~1), its size (argument~2), its destination rank (argument~4), and the new communicator (argument~6).
DMRlib provides the variable \texttt{DMR\_INTERCOMM} to represent the inter-communicator.

The child processes invoke the second function in the listing (line~9). These are responsible for receiving the data and continuing with the execution.
The scalability factor is again calculated using the same operation as in the former case (line~10). However, since the child processes executed this, the variable names are swapped.
With this \textit{factor}, we obtain the source rank of the data in the parent communicator (line 11).
The chunk size is calculated (line 12) to allocate memory for the data structure (line 13).
In this procedure, variables \textit{data\_size} and \textit{data} are overwritten; the data array (\textit{data}) is a null pointer in the memory of the child processes, and hence we have to allocate the necessary memory.
Eventually, the \texttt{MPI\_Recv} operation is invoked to retrieve the memory pointer to store the data (argument~1); the number of received elements (argument~2); the source rank in the initial communicator (argument~4); and the initial communicator itself (argument~6).

\begin{lstlisting}[float,caption=User data redistribution functions for an expansion., label=code:dmrlib-user-expansion, captionpos=b]
void send_expand(double *data, int data_size) {
  factor = dmr_intercomm_nprocs / comm_world_nprocs;
  new_data_size = data_size / factor;
  for (i = 0; i < factor; i++) {
    dst_rank = my_rank * factor + i;
    MPI_Send(data + new_data_size * i, new_data_size, MPI_DOUBLE, dst_rank, tag, DMR_INTERCOMM);
  }
}
void recv_expand(double **data, int *data_size) {
    factor = comm_world_nprocs / dmr_intercomm_nprocs;
    src_rank = my_rank / factor;
    *data_size = (*data_size) / factor;
    *data = malloc((*data_size) * sizeof (double));
    MPI_Recv(*data, *data_size, MPI_DOUBLE, src_rank, tag, DMR_INTERCOMM, MPI_STATUS_IGNORE);
}
\end{lstlisting}

\subsection{Predefined Redistribution Patterns}
\label{subsec:patterns}
Although the data redistribution can be performed via standard MPI clauses, in an effort to ease the coding of that procedure, we provide two sets of predefined functions that accommodate some common communication patterns:

\begin{itemize}
\item \textbf{Default Redistribution}: classic \reviewR{1D} uniform distribution to a number of processes multiple of or divisible by the original number (e.g., Figure~\ref{fig:expand}).
\item \textbf{Block Cyclic Redistribution}: implements the data
  transfers necessary when \reviewR{1D} data is block-cyclically distributed over
  the processes\footnote{\url{https://computing.llnl.gov/tutorials/parallel_comp/#distributions}}.
\end{itemize}
DMRlib can be expanded with other communication patterns depending on the application necessities.

Table~\ref{tab:dmrlib-predef} shows the headers of the redistribution functions.
All of them receive the same two initial arguments: the pointer to the data (when receiving, a pointer address) and the MPI datatype.
For the default redistribution, they also receive the number of elements of the data array;
while for the block-cyclic pattern, the functions receive the number of blocks and their size.

\begin{table}
\caption{Predefined Redistribution Headers in DMRlib}\label{tab:dmrlib-predef}
\ra{0}
\centering\resizebox{\linewidth}{!}{%
\begin{tabular}{l}\toprule
\textbf{Default Redistribution}\\\midrule
void DMR\_Send\_expand\_default(void *data, MPI\_Datatype type, int size);\\
void DMR\_Recv\_expand\_default(void **data, MPI\_Datatype type, int *size);\\
void DMR\_Send\_shrink\_default(void *data, MPI\_Datatype type, int size);\\
void DMR\_Recv\_shrink\_default(void **data, MPI\_Datatype type, int *size);\\\midrule
\textbf{Block Cyclic Redistribution}\\\midrule
void DMR\_Send\_expand\_blockcyclic(void *data, MPI\_Datatype type, int n, int size);\\
void DMR\_Recv\_expand\_blockcyclic(void **data, MPI\_Datatype type, int *n, int *size);\\
void DMR\_Send\_shrink\_blockcyclic(void *data, MPI\_Datatype type, int n, int size);\\
void DMR\_Recv\_shrink\_blockcyclic(void **data, MPI\_Datatype type, int *n, int *size);\\
\bottomrule
\end{tabular}}
\end{table}

In the example in Listing~\ref{code:dmrlib-usage}, we could have implemented the same functionality using the default redistribution pattern as presented in Listing~\ref{code:dmrlib-pattern}, hence saving the user from implementing the data redistribution presented in Listing~\ref{code:dmrlib-user-expansion}.
\begin{lstlisting}[float,caption=DMRlib malleability with the default redistribution functions., label=code:dmrlib-pattern, captionpos=b]
void compute(double *data, int data_size, int step) {
  DMR_Set_parameters(min, max, pref);
  for (int i = step; i < TOTAL_STEPS; i++) {
    DMR_RECONFIG(compute(data, data_size, i), DMR_Send_expand_default(data, MPI_DOUBLE, data_size), DMR_Recv_expand_default(&data, MPI_DOUBLE, &data_size), DMR_Send_shrink_default(data, MPI_DOUBLE, data_size), DMR_Recv_shrink_default(&data, MPI_DOUBLE, &data_size));
    /* Computation */
  }
}
\end{lstlisting}
\review{
Further details about more complex data redistribution using DMRlib can be found in~\cite{Iserte2019}, and ~\cite{Iserte2018hpg}}

\section{Usability Study}\label{sec:usability}
This section studies the usability of the frameworks introduced in Section~\ref{sec:related}, from the point of view of their semantics and characteristics.
This study aims to compare how the different malleability solutions implement the same minimal use case. 
The comparison continues with an evaluation of their features, and a discussion of their potential adoption in production environments.
The section is closed with a description of how malleability is implemented with DMRlib in a set of applications, executed afterward, in Section~\ref{sec:perf_eval}.

As the works based on SCR extensions~\cite{Lemarinier2016}, EasyGrid AMS~\cite{Ribeiro2013}, ULFM~\cite{Lemarinier2016}, and ReSHAPE~\cite{Sudarsan2007}, do not publicly provide any example on how to implement malleability,
we limit the study to PCM ~\cite{ElMaghraoui2007}, AMPI~\cite{Gupta}, Flex-MPI~\cite{Martin2013}, Elastic MPI~\cite{Compres2016} and DMR~\cite{Iserte2017}.

\subsection{Usability Analysis}
Malleability usually targets iterative applications, which contain a main loop representing an ideal synchronization point (or malleability point) in the code for re-scaling.
In order to showcase malleability using the different frameworks, we consider a prototype iterative application that performs calculations over a single data array called \textit{data}, of size determined by a \textit{dataSize} variable.
Using this generic application, we perform a \emph{migration} using pseudo--MPI code and the tools offered by each malleability solution.
\review{
We note that \textit{migration} is the simplest case of reconfiguration because the number of processes does not change.
In this regard, a migration will suffice to illustrate and compare the main features of the different tools.
In a practical scenario, reconfigurations would mainly involve job expansions or shrinkages. Subsection~\ref{subsec:handson} describes how expansions are implemented with DMRlib.
}

Listing~\ref{code:use-mpi} contains the skeleton of the application mentioned above using a pure MPI approach.
All the examples in this section share the same skeleton: the \textit{main} function initializes the data and then calls the \textit{compute} function.
This function contains the main loop, where the actual computation occurs and where the malleability is realized.
Specifically, in this code, we check whether there is a parent communicator (line~4), which would indicate that the processes are in the middle of a reconfiguration.
If that is not the case, the execution continues with the computation (line~12).
In case there is a parent communicator, the processes have to receive the data from their counterparts in the initial communicator (lines~8--10).

\begin{lstlisting}[float,caption=Pseudo-code of a migration using bare MPI., label=code:use-mpi, captionpos=b]
void main(int argc, char **argv) {
  MPI_Init(&argc, &argv);
  MPI_Comm_get_parent(&parentComm);
  if (parentComm == MPI_COMM_NULL) {
    step = 0;
    /* Initialization */
  } else {
    MPI_Recv(&dataSize, myRank, parentComm);
    MPI_Recv(data, myRank, parentComm);
    MPI_Recv(&step, myRank, parentComm);
  }
  compute(data, dataSize, step);
  MPI_Finalize();
}
void compute(double *data, int dataSize, int step) {
  for (t = step; t < TIMESTEPS; t++) {
    nodeList = get_new_nodelist_somehow();
    if (nodelist != NULL) {
      MPI_Comm_spawn(myapp.bin, nodeList, &newComm);
      MPI_Send(dataSize, myRank, newComm);
      MPI_Send(data, myRank, newComm);
      MPI_Send(t, myRank, newComm);
      MPI_Finalize();
      exit(0);
    }
    /* Computation */
  }
}
\end{lstlisting}

In the computation stage (line~16), for each step, we use a function that represents a call to a scheduler that returns a list of nodes (line~17).
If the list is void, the computation continues normally (line~26).
Otherwise, a reconfiguration has to be performed.
In such a case, the new processes are spawned (line~19), and the data is sent to them (lines 20--22).
Once the data is received, the initial processes terminate their
execution (line~24), and their recently-created counterparts continue with the execution.

\subsubsection{PCM}

\begin{lstlisting}[float,caption=Pseudo-code of a migration using the PCM API., label=code:use-pcm, captionpos=b]
void main(int argc, char **argv) {
  PCM_MPI_Init(&argc, &argv);
  PCM_COMM_WORLD = MPI_COMM_WORLD;
  PCM_Init(PCM_COMM_WORLD);
  PCM_Status status = PCM_Process_status;
  if (status == PCM_STARTED) {
    step = 0;
    /* Initialization */
  } else {
    PCM_Load(myRank, "step", &step);
    PCM_Load(myRank, "dataSize", &dataSize);
    PCM_Load(myRank, "data", data);
  }
  compute(data, dataSize, step);
}
void compute(double *data, int dataSize, int step) {
  for (t = step; t < TIMESTEPS; t++) {
    pcm_status = PCM_Status(PCM_COMM_WORLD);
    if (pcm_status == PCM_MIGRATE) {
      PCM_Store(myRank, "step", &t, PCM_INT, 1);
      PCM_Store(myRank, "dataSize", &dataSize, PCM_INT, 1);
      PCM_Store(myRank, "data", data, PCM_DOUBLE, dataSize);
      PCM_COMM_WORLD = PCM_Reconfigure(PCM_COMM_WORLD, argv[0]);
    } else if (pcm_status == PCM_RECONFIGURE) {
      PCM_Reconfigure(&PCM_COMM_WORLD,argv[0]); 
      MPI_Comm_rank(PCM_COMM_WORLD, &rank);
    }
    /* Computation */
  }
}
\end{lstlisting}

Following the example presented in~\cite{ElMaghraoui2009}, we have adopted malleability in our skeleton code using the PCM API (Listing~\ref{code:use-pcm}).
PCM wraps many of the MPI functions/variables, but the workflow mimics that presented in Listing~\ref{code:use-mpi}.
In line~6, we check the \textit{status}, and provided there is a migration in progress, the data is loaded in lines~10--12. The \textit{compute} function (line~16) also presents many similarities with the pure MPI implementation.
\review{
The function ``PCM\_Status'', in line~18, assesses which reconfiguration action has been scheduled.
Although we only consider migration, the PCM API provides further reconfiguration actions. 
As the new processes have to load the previous user data, initial processes must keep the data (lines~20--22).
Finally, the reconfiguration is concluded with a call to \textit{PCM\_Reconfigure}. This is a collective function that needs to be called by both the migrating (line~23) and non-migrating processes (line~25). 
}

\subsubsection{AMPI}
AMPI provides automated support, via CHARM++, for migrating MPI ranks among nodes.
\optional{
in a system without any application-specific code\footnote{\url{https://charm.readthedocs.io/en/latest/ampi/manual.html}}.}
\review{
CHARM++ is based on migratable objects called \textit{chares}, which are virtualized processes associated with user-level threads. For this reason, these virtualized objects are easy to send/receive from one host to another.
}
In the example shown in Listing~\ref{code:use-ampi}, we have used
``isomalloc''\footnote{\url{https://charm.readthedocs.io/en/latest/tcharm/manual.html#migration-based-load-balancing}}, which allows other worker threads in the system to allocate slices of virtual memory for all user-level threads, enabling transparent migration of memory pointers.
AMPI also provides data registration mechanisms as alternate allocation procedures as well as tools for data pack\slash unpack in order to perform data redistribution.

\begin{lstlisting}[float,caption=Pseudo-code of a migration using AMPI., label=code:use-ampi, captionpos=b]
void main(int argc, char **argv) {
  step = 0;
  /* Initialization */
  compute(data, dataSize, step);
}
void compute(double *data, int dataSize, int step) {
  MPI_Info_create(&hints);
  MPI_Info_set(hints, "ampi_load_balance", "sync");
  for (t = step; t < TIMESTEPS; t++) {
    /* Computation */
    AMPI_Migrate(hints);
  }
}
\end{lstlisting}

Since we assume implicit registration of data provided by ``isomalloc'' during initialization (line~3), the \textit{main} function initiates the computation in line~4.
The reconfiguration is set using an ``MPI\_Info'' object (lines~7--8), and the iterations are initiated.
\review{
The hint \textit{ampi\_load\_balance==sync} reports to CHARM++ that the application is already at a synchronization point, and the data redistribution will be performed synchronously.
}
For each iteration, the data is computed (line~10), and the migration is invoked (line~11).

\subsubsection{Flex-MPI}
Flex-MPI is not only a malleability solution but also a performance-aware framework to monitor the execution performance in each iteration.
Flex-MPI can schedule the most appropriate reconfiguration action, which leads to a resultant code presenting a higher level of instrumentation.

Based on~\cite{martin2015}, we have tailored Listing~\ref{code:use-flex} to show the malleable implementation of our sample code using Flex-MPI.
At the beginning of the program, data is initialized and registered prior to the computation stage (lines~4--9).
Then, in each computational step (line~13), the performance monitor is initiated, and the computation is performed (lines~14--15).
A reconfiguration action is executed with the gathered information during the step execution (line~16).
After a reconfiguration, unlike the previous malleability solutions, the execution flow does not return to the \textit{main} function, but remains in \textit{compute} for the remainder of the run.
Although this study is focused on the migration action, Flex-MPI implements a simple procedure for removing processes in case of shrinkage.
Lines~18--19 check each process and terminate those processes selected by the runtime.

\begin{lstlisting}[float,caption=Pseudo-code of a migration using Flex-MPI., label=code:use-flex, captionpos=b]
void main(int argc, char **argv) {
  MPI_Init(&argc, &argv);
  step = 0;
  /* Initialization */
  XMPI_Get_wsize();
  XMPI_Register(dataSize);
  XMPI_Register(data);
  XMPI_Register(step);
  XMPI_Get_Shared_data();
  compute(data, dataSize, step);
}
void compute(double *data, int dataSize, int step) {
  for (t = step; t < TIMESTEPS; t++) {
  	XMPI_Monitor_init();
    /* Computation */
    XMPI_Eval_reconfiguration();
    status = XMPI_Get_process_status();
    if (status == XMPI_REMOVED)
      break;
  }
}
\end{lstlisting}

\subsubsection{Elastic MPI}
Listing~\ref{code:use-mpich} shows a tentative implementation of a migration using Elastic MPI, following the malleable implementation of a producer--consumer scheme in~\cite{Compres2016}.
\optional{
Although Elastic MPI does not support a complete migration of processes (the node where \texttt{srun} is executed cannot be substituted because of an intrinsic limitation of the model), our pseudo-code assumes that this constraint is overcome.
}
\begin{lstlisting}[float,caption=Pseudo-code of a migration using Elastic MPI., label=code:use-mpich, captionpos=b]
void main(int argc, char **argv) {
  MPI_Init_adapt(&argc, &argv, &status);
  if (status == JOINING) {
    MPI_Probe_adapt(&adapt);
    if (adapt == ADAPT_TRUE) {
      MPI_Comm_adapt_begin();
      /* Data redistribution code */
      MPI_Comm_adapt_commit();
    }
  } else {
    if (status == NEW) {
      step = 0;
      /* Initialization */
    }
  }
  compute(data, dataSize, step, status);
}
void compute(double *data, int dataSize, int step, int status) {
  for (t = step; t < TIMESTEPS; t++) {
    MPI_Probe_adapt(&adapt);
    if ((status == JOINING) || (adapt == ADAPT_TRUE)) {
      MPI_Comm_adapt_begin();
      /* Data redistribution code */
      MPI_Comm_adapt_commit();
    }
    /* Computation */
  }
}
\end{lstlisting}

The program starts with a tuned version of \texttt{MPI\_Init}, which accepts a new parameter for malleability (line~2).
This new parameter specifies whether the process has been created by Slurm (line~3).
If this is the case, the process tests whether it has to be adapted to a new process layout (line~4) in order to perform reconfiguration and data redistribution (lines~6--8).
The authors of this work do not provide sufficient information on how to perform data redistribution; however, they do specify where it occurs (lines~7 and~23).
If the process is created by the launcher (line~11), the application performs the original initialization of variables.

Once the program is initialized, the execution continues in the \textit{compute} function (line~18).
For each iteration, the processes probe if an adaption is ongoing (line~20). If so, reconfiguration and redistribution
are performed (lines~22--24).
Last, the processes execute their operations in line~26.

\subsubsection{DMR~API}
This project leverages the off-loading semantics of the OmpSs programming model~\cite{Sainz2015}, and hence the reconfiguration is driven by \textit{\#pragma} directives.

Listing~\ref{code:use-dmr} presents an implementation of malleability with the DMR~API.
The original \textit{main} function is unaltered, initializing the data and invoking the \textit{compute} function.
Once the iterations start (line~8), a call to the DMR~API is performed (line~9).
This call returns the reconfiguration action scheduled by the RMS and the MPI communicator where the new processes are spawned.
The \textit{\#pragma} in line~11 conducts the reconfiguration. At this point, the user specifies the data dependencies and the communication pattern among the processes between communicators.
This directive explicitly states that \textit{data} is an input dependency for the new processes in the \textit{handler} MPI communicator.
Since the goal is to implement a migration, the communication among ranks is \textit{one--to--one}.
Furthermore, in order to continue the program in the same function, instead of returning to the \textit{main} function after the \textit{\#pragma}, the user indicates the resuming point for the execution (line~12).
The initial processes are automatically terminated by the runtime after the task is successfully off-loaded to the new processes.

\begin{lstlisting}[float,caption=Pseudo-code of a migration using the DMR~API., label=code:use-dmr, captionpos=b]
void main(int argc, char **argv) {
  MPI_Init(&argc, &argv);
  step = 0;
  /* Initialization */
  compute(data, dataSize, step);
}
void compute(double *data, int dataSize, int step) {
  for (t = step; t < TIMESTEPS; t++) {
    action = dmr_check_status(&handler);
    if (action == MIGRATION) {
      #pragma omp task in(data) onto(handler, myRank)
      compute(data, dataSize, step);
    } else {
      /* Computation */
    }
  }
}
\end{lstlisting}

\subsection{Usability Evaluation}
\review{
Notwithstanding the variety of methods to adopt malleability in parallel scientific applications, none of them combines all the features that we consider crucial in order to gain popularity among developers:
\begin{inparaenum}[i)]
  \item automatic support for data transfers in job reconfiguration; and
  \item MPI-like syntax based on the MPI standard without dependencies on any particular MPI library.
\end{inparaenum}
\newline\indent
Compared with the reviewed approaches, the solution presented in this paper, DMRlib, improves the appeal and the state-of-the-art of process malleability, by providing a simple trigger mechanism for reconfiguration that hides all the reconfiguration internals and performs the data redistribution among the processes through \textit{standard MPI routines}.
This section evaluates the usability of DMRlib compared to the previously studied malleability solutions.
}

Table~\ref{tab:tools} compares the usability features of the bare MPI with the malleability frameworks studied in Section (first column).
\review{
The second column corresponds to the Source Lines of Code (SLOC) metric.
}
\optional{SLOC has nothing to do with productivity, but it is one of the most efficient ways to compare applications that have been developed within the same context~\cite{McConnell2006}.
}
\review{
The values of SLOC for each framework in Section~\ref{sec:usability} correlates respectively to Listings~\ref{code:use-mpi}--\ref{code:use-dmr}, and Listing~\ref{code:dmrlib-usage} for DMRlib (adding the lines corresponding to the \textit{main} function).
}
For Elastic MPI, the number of lines does not include the data redistribution code.

\begin{table}
    \caption{Malleability solutions usability features comparison}
    \label{tab:tools}
    \centering
    \resizebox{\linewidth}{!}{%
    \begin{tabular}{r|ccccc}
    \toprule
    & SLOC & Data transfer & Standard MPI & Paradigm & RMS\\ \midrule
    Bare MPI & 28 & Manual & Yes  & MPI & -\\
    PCM API & 30 & Manual & No/MPICH & MPI & -\\
    AMPI & 13 & Auto* & Yes & CHARM++ & Torque/Maui\\
    Flex-MPI & 21 & Auto & No/MPICH  & MPI & -\\
    Elastic MPI & 26* & Manual & No/MPICH  & MPI & Slurm\\
    DMR~API & 17 & Auto & Yes & OmpSs & Slurm\\
    DMRlib & 13 & Auto* & Yes  & MPI & Slurm\\
    \bottomrule
    \end{tabular}}
\end{table}

The reduction of the SLOC is closely related to the type of data transfers (third column).
We conclude that solutions with any type of automatic data transfers (AMPI, Flex-MPI, DMR~API, DMRlib) drastically reduce the coding effort since the task is off-loaded to the runtime.
Notably, we catalog AMPI and DMRlib as \textit{auto*}, since the provided mechanisms (``isomalloc'' in AMPI and the \textit{predefined redistribution patterns} in DMRlib), which can be used to perform automatic data redistributions in many cases. In contrast, Flex-MPI and the DMR~API, employing \textit{data structure registers} and \textit{data dependencies} respectively, provide automatic data redistributions among processes in every case.
 
An important additional issue is the solution dependency on the underlying MPI library.
The fourth column of Table~\ref{tab:tools} indicates whether the frameworks can operate over any MPI-2 Standard library or if they are based on a specific MPI implementation.
Honoring the standard is relevant because those solutions can be expected to be more reliable, portable, and long-lasting.

In terms of usability, we consider the most relevant: (i) a lower number of lines, (ii) automatic data transfers, (iii) support for any MPI standard implementation, (iv) the fact that the framework does not modify the MPI standard, and (v) its integration in an RMS.
Although AMPI and the DMR~API meet all those requirements, we also consider crucial that the solution relies on the MPI paradigm instead of CHARM++ or OmpSs.
The majority of solutions adhere to the MPI programming paradigm (fifth column) as it is widely accepted in HPC.
Finally, the last column specifies whether the solution is integrated into an RMS.

All in all, we can establish DMRlib as the solution to the highest appeal for malleability.

\subsection{DMRlib Hands-on Experience}\label{subsec:handson}
We have leveraged DMRlib to develop malleable implementations of the Conjugate Gradient (CG)
solver~\cite{Hestenes1952}, the Jacobi
method~\cite{Saad2003}, the N-body
problem~\cite{Aarseth2009}, and the bioinformatics HPG-aligner tool~\cite{Medina2016}.

For CG and Jacobi, the coding process for malleability is quite similar, although these two methods do not feature the same number/type of data structures.
In particular, Jacobi operates a flat-stored square matrix plus two arrays, while CG handles two additional arrays.
For CG, the \texttt{DMR\_RECONFIG} call may be invoked as follows:

\begin{Verbatim}[fontsize=\scriptsize]
DMR_RECONFIG( CG(m, a1, a2, a3, a4, size, step),
  send_expand(m, a1, a2, a3, a4, size),
  recv_expand(&m, &a1, &a2, &a3, &a4, &size),
  send_shrink(m, a1, a2, a3, a4, size),
  recv_shrink(&m, &a1, &a2, &a3, &a4, &size));
\end{Verbatim}

For instance, the sending function for an expansion may include the following calls to the redistribution functions:
\begin{Verbatim}[fontsize=\scriptsize]
void send_expand(double *m, double *a1, ..., int size){
  DMR_Send_expand_default(m, MPI_DOUBLE, size * size);
  DMR_Send_expand_default(a1, MPI_DOUBLE, size);
  ...}
\end{Verbatim}

N-body only handles an array in the redistribution.
However, this structure is composed of \textit{particles}, which is a non-standard data type.
\review{
For this reason, we created a new MPI datatype (named \texttt{MPI\_PARTICLE}) composed of two 3D vectors, one for the particle position and the other velocity, and two floats for the mass and weight.
}
With this datatype, we leverage the predefined redistribution functions as follows:

\begin{Verbatim}[fontsize=\scriptsize]
DMR_RECONFIG( N-body(particles, size, step),
  DMR_Send_expand_default(particles, MPI_PARTICLE, size),
  DMR_Recv_expand_default(&particles, MPI_PARTICLE, &size),
  DMR_Send_shrink_default(particles, MPI_PARTICLE, size),
  DMR_Recv_shrink_default(&particles, MPI_PARTICLE, &size));
\end{Verbatim}

Finally, for HPG-aligner, we developed ad-hoc redistribution functions, since it does not present a regular communication pattern~\cite{Iserte2018hpg}.
Also, HPG-aligner features a producer--consumer architecture, where two processes are in charge of reading/writing data while the remaining processes act as \textit{workers}.
For this reason, the minimum number of processes required to run this application is three (at least it needs one \textit{worker} plus reader and writer processes).

The processes in applications such as CG, Jacobi, or N-body work over data subsets exchanged at each iteration.
However, the ranks in HPG-aligner work-on independent chunks of data, writing the results to disk after computing each chunk.
\review{
This behavior makes HPG-aligner an I/O intensive application with limited scalability.
\newline\indent
This set of applications is designed to cover different scalability behaviors and to be representative of a large variety of applications.
}
\optional{
from N-body, which is expected to remain highly relevant during the next decade~\cite{AghababaieBeni2017}, to applications in various domains such as scientific computing simulations, as well as a producer--consumer application for bioinformatics like HPG-aligner.
}

\section{Performance Evaluation}\label{sec:perf_eval}
In this section, we evaluate the benefits of malleable workloads composed of the previously described applications in a production environment. 
Our evaluation was performed using \textbf{129 nodes} of the Marenostrum IV supercomputer at the Barcelona Supercomputing Center (BSC). 
Each node in this facility is equipped with 2 Intel Xeon Platinum~8160 sockets (24 cores at 2.10~GHz each) for a total of 48 cores with 96~GB of RAM.
The nodes are interconnected through a 100~Gbit/s Intel Omni-Path network.
For the software stack, we used MPICH 3.2, OmpSs 15.06, and Slurm 15.08.
Slurm was configured with the following plug-ins:
\begin{itemize}
\item Job scheduling: \texttt{sched/backfill} with a 10-second interval time among scheduling attempts.
\item Job priority: \texttt{priority/multifactor} without defining a wall time request for jobs.
\item Resource selection: \texttt{select/linear}.
\end{itemize}
One of the nodes hosted the Slurm management daemon, while the remaining ones were used as compute nodes.

\subsection{Job Malleability Scheduling Policy}\label{subsec:sched}
Applications are submitted as jobs to the workload manager.
In~\cite{Lublin2003}, the authors categorize jobs depending on the user participation when determining the number of processes on initialization or during the execution. 
The most interesting job categories for system-aware reconfiguration are:
\begin{itemize}
    \item Rigid job: initiated with a static number of processes defined by the user.
    \item Moldable job: the RMS decides the number of processes on initialization.
    \item Malleable job: the number of processes can be changed during the execution without user intervention.
\end{itemize}
On this basis, we have classified jobs combining moldability and malleability.
Table~\ref{tab:new-classification} shows this new classification, distinguishing whether jobs can be resized on initialization or during the execution.
For example, if a non-malleable job is rigid, the job will be referred to as ``fixed''.
In contrast, if a malleable job is moldable, we consider it to be ``flexible''.

\begin{table}
\caption{Job classification depending how it can be resized} \label{tab:new-classification}
\scriptsize
\centering
\begin{tabular}{c||c|c}\toprule
\textbf{Job Malleable?} & \textbf{Rigid submission} & \textbf{Moldable submission}\\ \midrule
No & Fixed & Pure Moldable\\
Yes & Pure Malleable & Flexible\\ 
\bottomrule
\end{tabular}
\end{table}

In this work, we rely on Slurm to schedule the jobs and
manage the resources. Apart from the rigid job submission,
where jobs request a fixed number of resources, Slurm
provides a moldable submission mechanism where a job may request
a range of resources.

We have implemented the reconfiguration policy for Slurm outlined in Algorithm~\ref{alg:slurm}.
This policy implements actions depending on the following malleability parameters:
\review{
\begin{itemize}
    \item Lower limit: the minimum number of processes.
    \item Upper limit: the maximum number of processes.
    \item Preferred: the number of processes determined by the user (usually based on a heuristic metric.)
\end{itemize}
}

When a job triggers a reconfiguration, Slurm first checks if the job is running with a number of resources lower than its \textit{preferred} configuration (line~1)---this can only happen in moldable submissions.
If this is the case, and there are available resources, the job is expanded (line~2) without exceeding the \textit{upper limit}.

The policy is designed to improve the throughput of the system.
Therefore, if there exist pending jobs in the queue (line~4), the policy checks if by shrinking a running job (but never with fewer processes than \textit{preferred}) and deallocating part of its resources (line~6), an additional job could be initiated.
When this action is scheduled for the first job, the additional one, responsible for the shrinking, is assigned the highest priority to run.
A job may only be shrunk if the number of resources allocated to it is higher than its \textit{preferred} value, and a pending job could benefit from the released resources;
otherwise, if no pending job can be initiated, and there are available resources, the job is expanded (line~9).
Finally, if the job is running with fewer processes than \textit{preferred}, and there are available resources, the job is expanded (line~11).
\begin{algorithm}[t]
  \scriptsize
\caption{Taking resize actions algorithm}\label{alg:slurm}
\begin{algorithmic}[1]
\If{$current < preferred$}
  \If{$avail\_resources$}
	\Return {expand}
\EndIf
\Else
\If{$pending\_jobs$}
	\If {$current > preferred$}
		\If {$\textit{an additional job can be initiated}$}
			\Return {shrink}
		\Else
			\If {$avail\_resources$}
					\Return {expand}
			\EndIf
		\EndIf
	\If {$avail\_resources$}
		\Return {expand}
	\EndIf
  \Else
  \If {$avail\_resources$}
			\Return {expand}
		\EndIf
	\EndIf
\EndIf
\EndIf
\end{algorithmic}
\end{algorithm}

\subsection{Applications Configuration}\label{subsec:apps}
The experiments in this paper involve the four applications used before.
Table~\ref{tab:apps-conf} shows the problem size and configuration of each application.
The application execution time is varied by adjusting the number of iterations instead of the problem size.
In order to evaluate the effect of malleability, we prioritize the number of iterations over the problem size to attain a reasonable execution time for a highly-utilized production supercomputer. 
Increasing the number of iterations accelerates the reaction time of the malleability system.
The results \reviewR{could be} extrapolated to larger problem sizes, where the iterations take longer.
\review{
All the jobs in this evaluation are executed with a single process per node and 48 threads per process.
}

\begin{table}
\caption{Applications configuration}
\label{tab:apps-conf}
\centering
\ra{0}
\resizebox{\linewidth}{!}{%
\begin{tabular}{rlc}
\toprule
Application &  Input Data & Iterations \\ \midrule
CG & A square matrix and 4 arrays of 32,768 elements & 10,000\\
Jacobi & A square matrix and 2 arrays of 16,384 elements & 10,000\\
N-body & 6,553,600 particles &  50\\
HPG-aligner & 40 millions 100-nucleotides reads & $\#workers \times 4$\\
\bottomrule
\end{tabular}}
\end{table}

\subsection{Applications Malleability Configuration}
In order to configure the malleability parameters of the applications (lower limit, preferred configuration, and upper limit), we performed a strong scalability test.
For this purpose, all the applications were executed with one process (except HPG-aligner, which was executed with three processes), and we doubled the number of processes incrementally at each step.
With those results, we obtain a relative factor of execution time decrease, referred to as \textit{gain difference}. 
The choice of the \textit{gain difference} is justified by its capacity to provide a fair balance between the number of resources allocated and the speedup obtained, as proven in several publications~\cite{Iserte2017},~\cite{Iserte2018hpg},~\cite{Iserte-thesis}.
\review{
The value of this heuristic decreases rapidly as the number of required nodes increases if there is not remarkable performance gain in between configurations. This heuristic rewards highly scalable applications with more resources for their malleability parameters.
}
The \textit{gain difference} is calculated as:
\[ s_{current} = \frac{t_{previous}-t_{current}}{t_{min\_procs}}\times 100\textnormal{,}\]
where $t_{current}$ is the completion time using the current number of processes; $t_{previous}$ is the completion time of the previous number of processes configuration; and $t_{min\_procs}$ is the completion time of the minimum number of processes configuration.
For example, to calculate the gain difference for HPG-aligner executed with 12 processes ($s(12)$), we subtract its time ($t(12)$) from the previous configuration completion time ($t(6)$).
This is divided by the reference completion time of the configuration with the lowest number of processes, in this case, $t(3)$ (the reference of the rest of the applications is $t(1)$). The result is finally multiplied by 100.

\begin{figure}
  \centering
  \includegraphics[clip, width=\columnwidth, trim={0cm 0cm 0cm 0cm}] {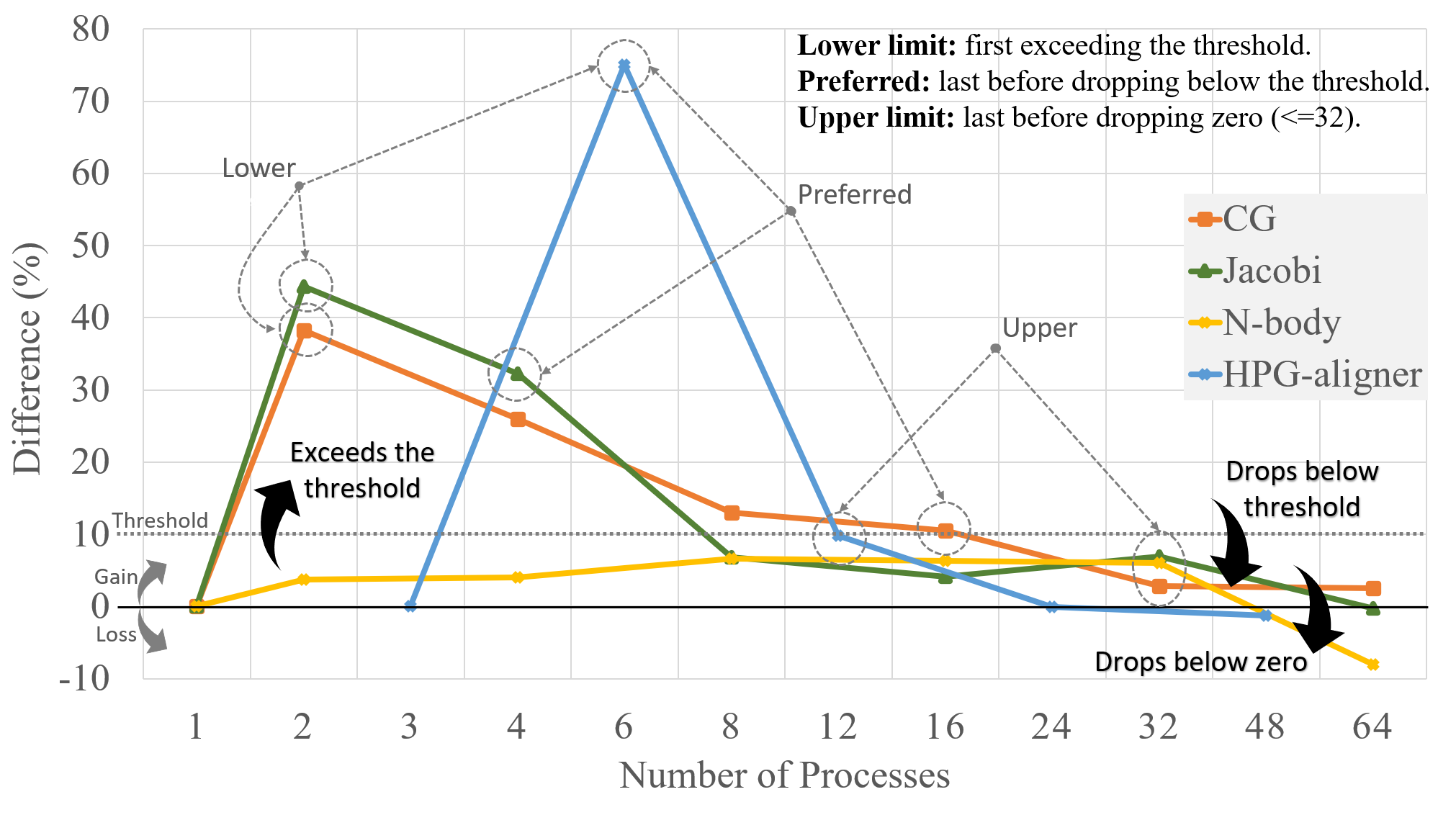}
  \caption{Gain difference of each application and a 10\% threshold (thick line) to determine the limits of malleability.}\label{fig:appsScal}
\end{figure}

Figure~\ref{fig:appsScal} depicts the gain difference for each configuration of all four applications.
\review{
With the \textit{gain difference} heuristic, enabling a 10\% threshold, the malleability parameters are defined as follows:
\begin{itemize}
    \item \textit{Lower limit}: first configuration exceeding the threshold.
    \item \textit{Preferred}: last configuration before dropping below the 10\% threshold. In other words, the configuration that delivers the fairest user-defined balance between performance and resources.
    \item \textit{Upper limit}: last configuration before dropping below zero (negative performance). In other words, the configuration that delivers the highest performance.
\end{itemize}
Note that applications that do not reach the threshold, such as N-body, have their lower limit and preferred value set to one process.
}

Although our cluster is composed of 128 compute nodes (plus one additional controller node), we restricted the jobs to request a maximum of 32 nodes, enforcing that a job does not monopolize more than a quarter of the cluster.

Table~\ref{tab:apps-mal} summarizes the results of the experiments with the malleable configurations.
The last column includes the scheduling inhibitor periods for CG and Jacobi. 
A 10-second period minimizes the number of reconfiguration requests reducing the overhead without a significant impact on the scheduling.
The remaining two applications do not need this inhibitor because their iterations are coarse-grained.

\begin{table}
\caption{Malleability parameters for the applications}
\label{tab:apps-mal}
\centering
\ra{0}
\resizebox{\linewidth}{!}{%
\begin{tabular}{rcccc}
\toprule
Application &  Lower Limit & Upper Limit & Preferred & Scheduling Period\\ \midrule
CG & 2 & 32 & 16 & 10 seconds\\
Jacobi & 2  & 32 & 4 & 10 seconds\\
N-body & 1 & 32 & 1 &  -\\
HPG-aligner & 6 & 12 & 6 & -\\
\bottomrule
\end{tabular}}
\end{table}

\subsection{Job Submission}
In Section~\ref{subsec:sched}, we introduced several job working modes, which distinguish whether the number of processes spawned by a job is determined before or during the execution.
For this purpose, the baseline of the following experiments is established by the results of the \textit{fixed} workload that submits rigid non-malleable jobs.
Furthermore, it is common that users submit their jobs with the configuration that provides the maximum performance (the \textit{upper limit}).

The moldable submission defines its range between \textit{lower} and \textit{upper} limits.
Table~\ref{tab:subs} describes how the jobs are submitted in both modes using the performance analysis of Table~\ref{tab:apps-mal}.

\begin{table}
\caption{Job submission in Slurm using \texttt{sbatch}}
\label{tab:subs}
\scriptsize
\ra{0}
\centering
\begin{tabular}{rll}
\toprule
Application &  Rigid Submission & Moldable Submission\\ \midrule
CG & -N32 ./cg & -N2-32 ./cg\\
Jacobi & -N32 ./jacobi & -N2-32 ./jacobi\\
N-body & -N32 ./nbody & -N1-32 ./nbody\\
HPG-aligner & -N12 ./hpgaligner & -N6-12 ./hpgaligner \\
\bottomrule
\end{tabular}
\end{table}

For all the tests, we have generated several workloads with jobs randomly corresponding to one of the four applications.
The workloads are composed of 100, 250, 500, 1,000, and 2,000 jobs, and feature the four different job versions: fixed, pure moldable, pure malleable, and flexible (see Table~\ref{tab:new-classification}).
With these sizes, the study aims to cover from small workloads, with hardly any pending job, to large workloads where the queue of pending jobs is significant.

The Feitelson model~\cite{Feitelson1996} determines the job inter-arrival time with a factor of 1, which represents a highly stressed scenario where jobs are massively submitted with a short delay that fits the Poisson distribution of the model.
\subsection{Experimental Results}\label{subsec:exp-res}
\begin{figure}
  \centering
  \includegraphics[clip,width=\columnwidth,trim={2.5cm 9cm 1.9cm 10.4cm}] {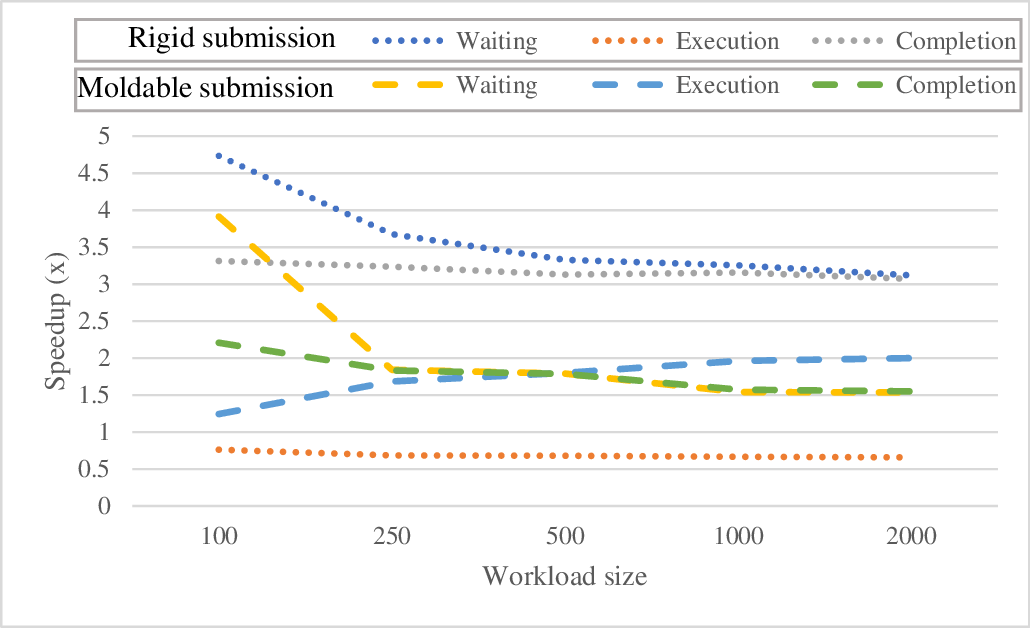}
  \caption{Comparison of the four types of workloads. The lines show the speedup
  attained thanks to the malleability for the average job waiting, execution, and completion time, grouped by submission mode.}\label{fig:speedups}
\end{figure}

Figure~\ref{fig:speedups} depicts the average job waiting, execution, and completion (waiting plus execution) times for each workload size.
The chart represents the speedup for the malleable workloads compared with their non-malleable counterparts.
Lines are grouped by submission mode: the dotted lines correspond to rigid submissions, while the thicker lines represent moldable submissions.

First, we analyze the rigid case (dotted lines).
Although the average job execution time increases for the malleable jobs ($speedup<1$), the completion time benefits from the reduction in the waiting time ($speedup~\simeq~3.25\times$).
The chart also shows a strong correlation between the completion time and the waiting time.
This leads to malleable jobs finalizing over 3x earlier than their non-malleable counterparts in a workload composed of jobs submitted in rigid mode.

In the case of moldable submissions (dashed lines), the speedup is more homogeneous once the workload reaches a minimum size.
We can observe the relevance of the waiting time for the job completion cost since the speedup lines almost overlap.
When the workload size increases, the queued jobs reach a saturation level where the waiting time cannot be improved further, and the speedup remains constant around 1.5x.
Apart from the waiting time, flexible jobs (moldable and malleable) show a higher speedup.
In a workload of pure moldable jobs (non-malleable submitted moldable), the execution time increases because jobs are likely to be initiated with fewer resources (it is easier to find a slot of 2 nodes rather than one of 32), and they have to finish their execution with their initial allocation.

\begin{figure}
  \centering
  \includegraphics[clip,width=0.85\columnwidth,trim={0cm 0cm 0cm 0cm}]{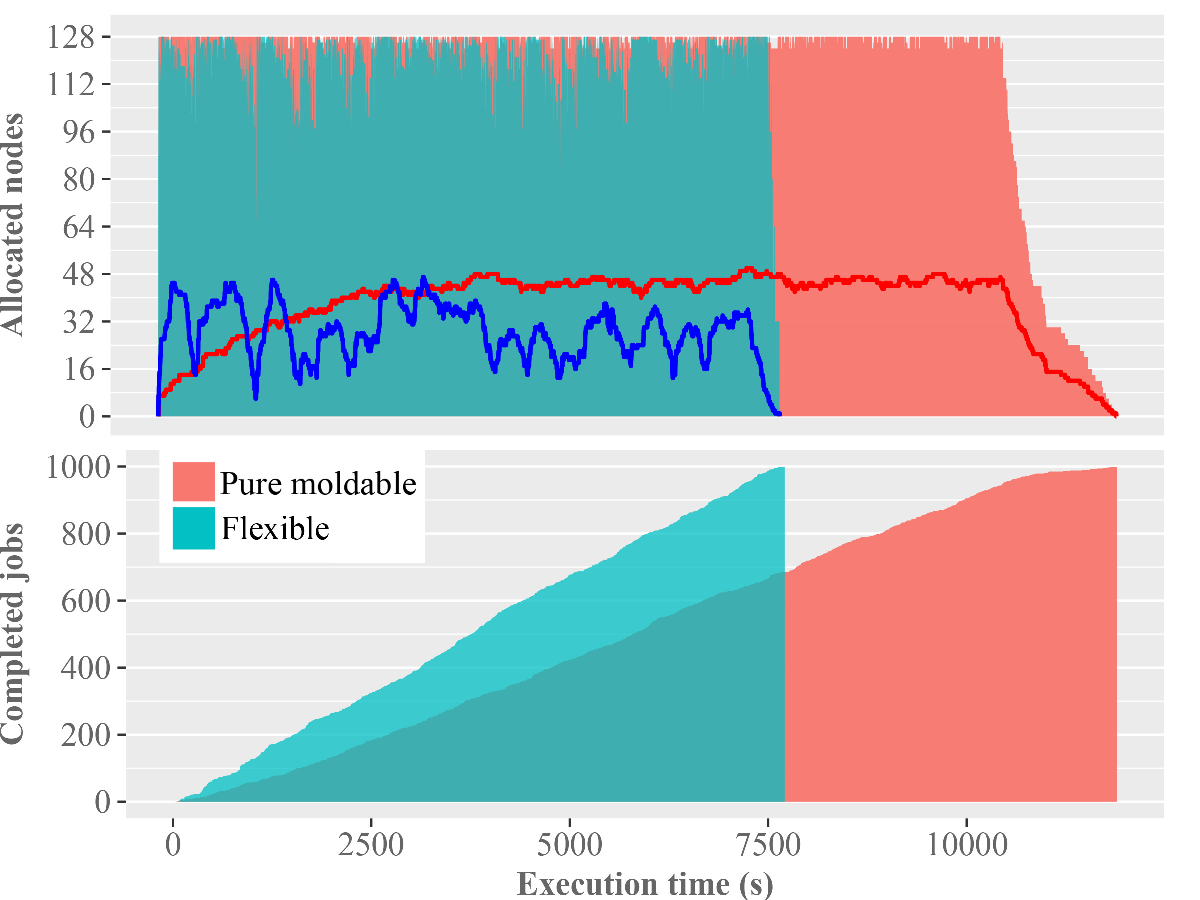}
  \caption{Comparison of the evolution in time of a 1,000-job workload for the
  pure moldable and flexible cases. The top chart represents the allocated resources (shapes) and the number of running jobs (lines). In the bottom chart, the shapes show the number of completed jobs in each second of the execution.}\label{fig:evo-mold-1000jobs}
\end{figure}

We next discuss in further detail the 1,000-job workload experiment with moldable submission.
Figure~\ref{fig:evo-mold-1000jobs} illustrates the behavior via the representation of the workload evolution over time.
In the top chart, the shapes report the evolution of the resource allocation over time, demonstrating that the flexible jobs can reallocate their resources to benefit from virtually all the nodes during the whole execution.
On the other hand, the pure moldable jobs maintain their initial allocation, resulting in a decrease in resource allocation near the second 10,000.
The lines in the top chart depict the evolution of the number of running jobs over time.
The pure moldable workload (red line) shows a regular evolution and a higher average number of running jobs since more jobs are running for a long time.
However, the flexible workload (blue line) redistributes the resources prioritizing the execution of a higher number of jobs, though with an allocation of nodes below the \textit{preferred} values.
The reconfiguration policy in Slurm always tries to avoid scenarios with a reduced load, as occurs at the beginning of the workload, initiating earlier new jobs.

The bottom chart in Figure~\ref{fig:evo-mold-1000jobs} represents the number of completed jobs across the timeline.
The green shape unveils a sharper increase in the throughput in terms of completed jobs per unit of time.
Accordingly to the speedup chart (Figure~\ref{fig:speedups}), the flexible jobs (malleable submitted moldable) complete their execution on average 1.5x faster than the non-malleable jobs when submitted as moldable (pure moldable).
This produces a 1.5x speedup of the global throughput when using malleability.
In the 100-job workload, the completion time speedup is 2.2x, and the workload is processed 3x faster than the pure moldable counterpart.

\begin{figure}
  \centering
  \includegraphics[clip,width=1.0\columnwidth,trim={0cm 0cm 0cm 0cm}] {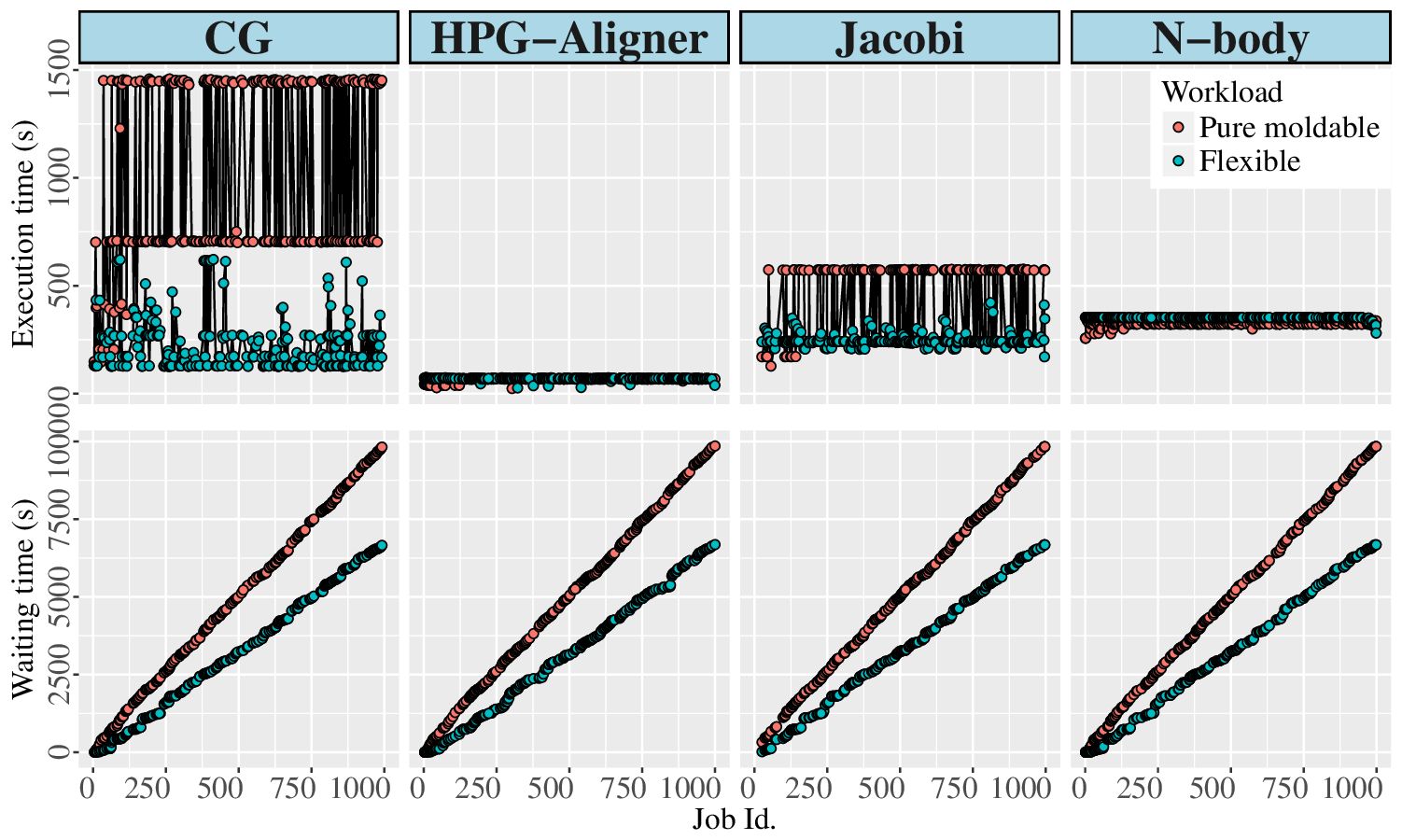}
  \caption{Execution and waiting times per job in the 1,000-job workload with moldable submission, grouped per application.}\label{fig:times-mold-1000jobs}
\end{figure}

Continuing with the same example, Figure~\ref{fig:times-mold-1000jobs} depicts the waiting and the execution times of each job comparing the pure moldable and the flexible versions.
In the top chart, the poorly-scalable applications (HPG-aligner and N-body) show virtually the same execution time in both versions.
However, applications like CG or Jacobi show higher variability in the execution time.
In the flexible case, expanding the jobs visibly tends to reduce the execution time.
Pure moldable jobs cannot be resized during their execution. As we have noted earlier, these are very likely to use a reduced set of resources, stretching their execution.
The bottom chart shows a regular behavior in the waiting time for all applications, showing an increase of more than 3,000 seconds for the last queued jobs.

\begin{figure}
  \centering
  \includegraphics[clip,width=1.0\columnwidth,trim={0cm 0cm 0cm 0cm}] {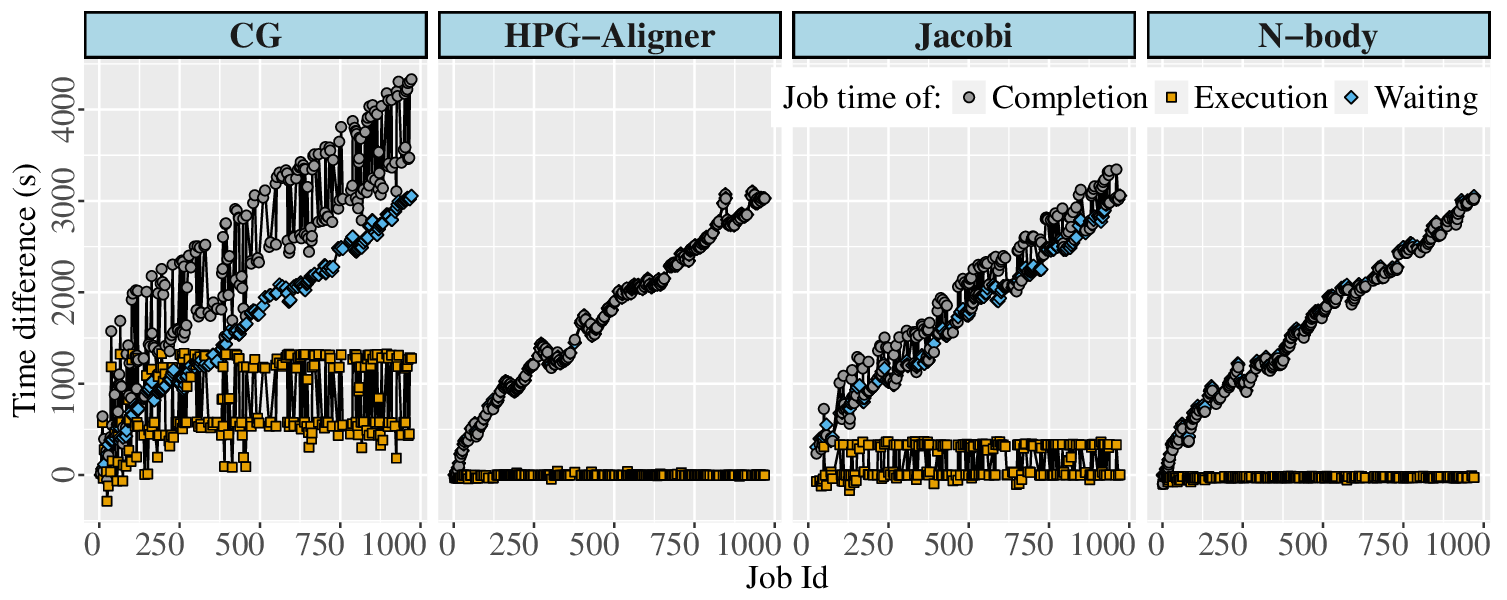}
  \caption{Time difference between the 1,000-job pure moldable and the flexible workloads, grouped per application} \label{fig:diff-mold-1000jobs}
\end{figure}

Figure~\ref{fig:diff-mold-1000jobs} gathers the waiting, execution, and completion times in the same chart and groups the time difference per application type for each job in both versions.
This chart reveals the strong correlation of the waiting time with the job completion time.

\begin{figure}[htp]
  \centering
  \subfloat[Workload completion time.]{%
  \includegraphics[clip,width=0.9\columnwidth,trim={1.9cm 10cm 1.85cm 11.38cm}] {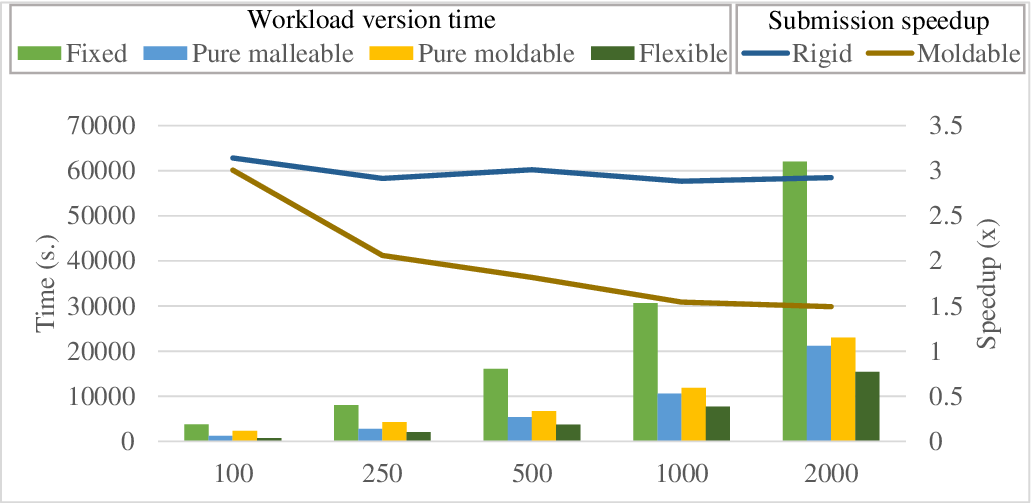}%
}\\
\subfloat[Average job execution time.]{%
\hspace{3mm}
  \includegraphics[clip,width=0.85\columnwidth,trim={2.4cm 10cm 2cm 11.6cm}] {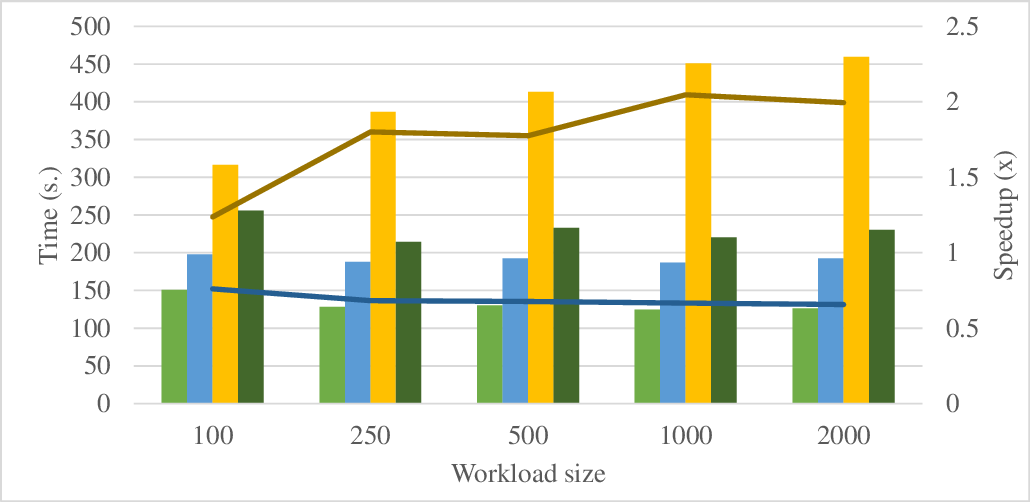}%
}
\caption{Workload type comparison and speedup of submission modes.} \label{fig:times}
\end{figure}

Figure~\ref{fig:times}a compares the workload completion time when the malleability is active in both submission modes: rigid and moldable.
The vertical bars represent the total completion time for each configuration.
Bars are grouped according to the workload size.
We have analyzed in detail the rigid submission (first and second bar of each group) and how the malleability improves it with speedups around 3x (blue line).
However, this chart also reveals the performance benefits of the moldable submission of non-malleable jobs (third bars).
We observe that, with the moldable submission, we obtain a similar completion time to that attained by a malleable workload with the traditional rigid submission.

These results suggest that the moldable submission may offer an easy--to--adopt high-throughput solution.
However, Figure~\ref{fig:times}b shows one of the most critical drawbacks for its adoption in production environments: the increase in job execution time.
While the individual average execution time for the rigid submission (first and second bars of each workload size) and flexible jobs (fourth bar) remain unaltered, in the pure moldable workloads (third bar), the jobs experience a notable growth of this metric.
To address this problem, we propose to leverage DMRlib to introduce job malleability.
\review{
Thanks to this, flexible workloads maintain the same time regardless of the workload size, unlike the pure moldable workload, yielding speedups of up to 2x in the average job execution time.
}

\begin{figure}
  \centering
  \includegraphics[clip,width=0.8\columnwidth,trim={2.5cm 5cm 2.5cm 5cm}] {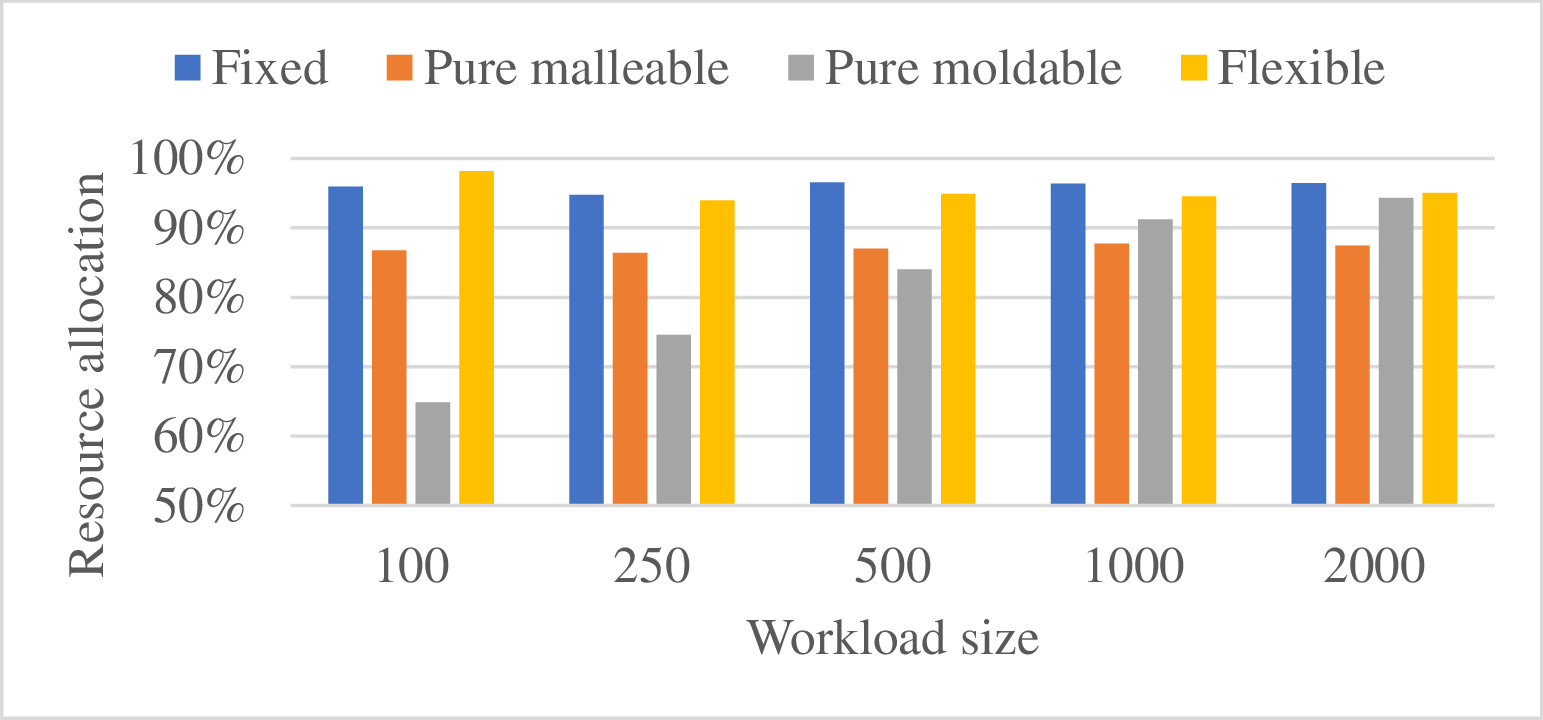}
  \caption{Workload type resource allocation comparison.}\label{fig:resources}
\end{figure}

An additional drawback of adopting the moldable submission without malleability is exposed in Figure~\ref{fig:resources},
which shows that the resource allocation rate drops for small workload sizes (third bar of each group).
In this case, the pure moldable workload under-utilizes the resources when the number of jobs in the queue is moderate.

We can consider flexible jobs, which solve these issues, as a remarkable technique for high-throughput computing (HTC). At the same time, DMRlib provides an easy--to--use approach.

\subsection{Impact of Malleability on the System}
In this section, we study heterogeneous workloads, where not all their jobs can be resized.
For this purpose, we have designed two types of experiments using the 1,000-job workload.
On the one hand, we have generated workloads with different percentages of malleable jobs,
specifically: 25\%, 50\% and 75\%.
On the other hand, we have created workloads where only one application is malleable. In other words, workloads where only one job type can be resized while the others remain fixed.
All the workloads feature two instances using either the rigid or the moldable submission.

Table~\ref{tab:notAllMal} contains the resource allocation rate and the percentage of time to execute the workload concerning the reference fixed workloads executions.
The reference values are placed in the yellow cells of the column labeled as ``None''.
In addition, we have also defined the results when all jobs are malleable (gray column labeled as ``All'') as a target reference.
\begin{table*}
\caption{Resource allocation and completion time of a 1,000-job workload with different rates of malleable jobs}
\label{tab:notAllMal}
\scriptsize
\centering
\ra{1}
\begin{tabular}{ R{1.4cm} R{1.55cm} C{1.05cm}  C{1cm} C{1cm} C{1cm}  C{1cm}  C{1.1cm} C{1.1cm} C{1.1cm} C{1.45cm} }
\toprule
 & & \multicolumn{9}{c}{Malleable Jobs}\\ \cmidrule{3-11}
Submission & Percentage & \cellcolor{yellow!75} None & 25\% & 50\% & 75\% & \cellcolor{gray!25} All & CG Only & Jacobi Only & N-body Only & HPG-aligner Only\\ \midrule
\multirow{2}{*}{Rigid} & Res. Alloc. & \cellcolor{yellow!75}96.37\% & 87.43\% & 87.07\% & 88.50\% & \cellcolor{gray!25}87.29\% & \cellcolor{blue!25}94.75\% & 88.34\% & 84.36\% & \cellcolor{blue!25}93.06\% \\
 & Comp. Time & \cellcolor{yellow!75}100.00\% & 92.52\% & 73.49\% & \cellcolor{green!75}53.77\% & \cellcolor{gray!25}36.12\% & 89.00\% & 105.67\% & \cellcolor{green!75}56.39\% & 101.77\% \\ \midrule
\multirow{2}{*}{Moldable} & Res. Alloc. & 91.23\% & 89.92\% & 88.97\% & 86.92\% & \cellcolor{gray!25}94.57\% & 95.51\% & 92.81\% & 90.62\% & 91.80\% \\
 & Comp. Time & 38.82\% & 37.29\% & 33.55\% & 30.09\% & \cellcolor{gray!25}25.15\%& \cellcolor{green!75}25.51\% & 43.44\% & 33.88\% & 38.40\% \\ \bottomrule
\end{tabular}
\end{table*}

\review{
In order to identify which kind of application has a greater impact on the results, the most interesting cells have been highlighted as follows:
\begin{itemize}
    \item Yellow is the reference when jobs are entirely static.
    \item Gray is the reference when all the jobs are malleable.
    \item Blue shows an allocation almost as the best case.
    \item Green highlights important reductions in the completion time when not all the jobs are malleable. 
\end{itemize}
}

For the rigid submissions (third and fourth rows), the heterogeneous workloads cannot beat the reference fixed workload resource allocation rate (96.37\%).
Nevertheless, those workloads where CG or HPG-aligner are malleable show quite a similar allocation rate.

Regarding the completion time, we have colored in green the 75\%-flexible and N-body-only workloads, which run in about half of the reference time (53.77\% and 56.39\%, respectively).
While the workloads with a 25-50-75\% malleable jobs define a progression where the execution time is inversely proportional to the rate of malleable jobs when only the N-body is malleable, the completion time is almost reduced to half (still far from the target reference (36.12\%)).

In the moldable submission case (fifth and sixth rows), the resource allocation rates are similar, since the moldable jobs can leverage the resources when launched.
However, the workload comprising malleable CG jobs can only achieve almost the same completion time (25.51\%) as that attained by the flexible workload (25.15\%).

From these results, we conclude that there is no correlation between resource allocation and execution time.
Besides, the resources are not wasted because the percentages fluctuate inside the reference rates.
In this study, we have also discovered that the rigid submissions profit more from poorly-scalable applications like our N-body.
For the moldable submissions, the resource manager can execute more efficiently those highly scalable applications such as our CG.
As shown in Figure~\ref{fig:appsScal}, CG increases the gain in performance up to 16 processes, being able to be shrunk up to two processes when needed.
Our malleability solution exploits this behavior efficiently.

\section{Conclusions}
\label{sec:conclusions}

In this paper, we have analyzed the process malleability state-of-the-art and how different frameworks offer distinct tools to accommodate malleability in the user codes.
Regarding the usability, we have postulated that a reduced number of lines, automatic data transfers, standard MPI-based, and the integration in a well-established HPC RMS, are the most desirable characteristics of an ideal malleability solution.

We have also presented a minimalist, MPI standard, data-transfer-aware malleability framework, named DMRlib, with support for the Slurm workload manager, 
This easy-to-adopt solution intends to facilitate turning a significant number of parallel scientific applications into malleable codes by setting reconfiguration points.

Using DMRlib, we have developed four malleable applications with different scalability patterns.
With these applications, we have generated fixed, pure moldable, pure malleable, and flexible workloads in order to assess the gains of malleability in terms of throughput, resource allocation, and energy consumption.

\optional{
An interesting study case would be if fixed jobs were submitted, requesting fewer resources.
In that scenario, the workload would probably present a higher throughput in benchmarks like those presented in this paper.
However, it would be less realistic since, generally, users submit their jobs at their maximum performance configuration, mainly because production workloads are not finite in time, and users cannot calculate the optimal job configuration that maximizes the system throughput.
For this reason, we kept this scenario out of the scope of this paper.
}

DMRlib has proven to reduce energy consumption while increasing the global throughput of HPC facilities.
In addition, our studies have demonstrated that it is not necessary to convert all the system applications into malleable; instead, by just adapting the right type of applications, the throughput can be dramatically boosted, and the energy consumption reduced more than 75\% (see Appendix~\ref{app:energy}).

DMRlib may implement arbitrary resize of processes (from any to any number of processes). 
However, in this paper, we have limited the reconfigurations to values multiple or divisible of the number of parent processes. 

The next natural step corresponds to the integration of DMRlib in intra-node malleability tools, such as dynamic load balancing (DLB)~\cite{Garcia-Gasulla2017}.

\section*{Acknowledgment}
This work was supported by projects TIN2014-53495-R, TIN2015-65316-P, and TIN2017-82972-R from MINECO and FEDER.
This project has received funding from the European Union's Horizon 2020 research and innovation programme under the Marie Sk\l{}odowska Curie grant agreement No. 749516.
Sergio Iserte was supported by a postdoctoral fellowship from Generalitat Valenciana and European Social Fund APOSTD/2020/026.
Finally, the authors want to thank the anonymous reviewers whose suggestions significantly improved the quality of this manuscript.

\bibliographystyle{IEEEtran}
\bibliography{paper-dmrlib}

\vspace{-1cm}
\begin{IEEEbiography}[{\includegraphics[width=1in,height=1.25in,clip,keepaspectratio]{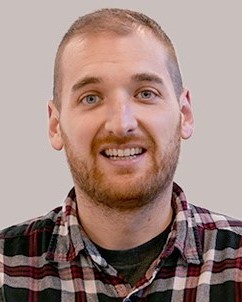}}]{Sergio Iserte}
holds the degrees of BS in Computer Engineering (2011), MS in Intelligent Systems (2014), and PhD in Computer Science (2018) from Universitat Jaume I, Spain.
He is currently postdoc researcher (APOSTD20) in the Mechanical and Engineering Construction Dept. at the same University.
He is currently involved in HPC projects related to parallel distributed computing, cloud computing, resource management, workload modeling, and deep learning for industrial applications.
\end{IEEEbiography}

\vspace{-1cm}
\begin{IEEEbiography}[{\includegraphics[width=1in,height=1.25in,clip,keepaspectratio]{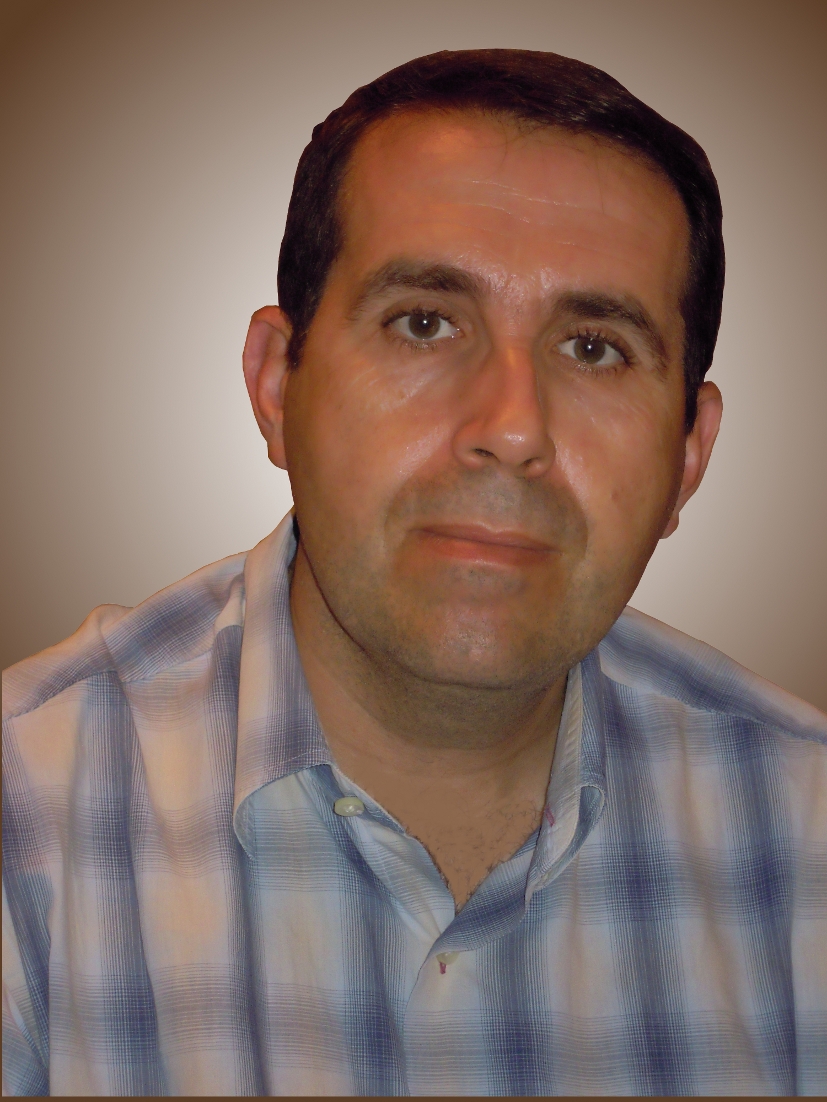}}]{Rafael Mayo}
 received the BS degree from Polytechnic
Valencia University in 1991. He obtained his PhD in Computer
Science in 2001 at the same University. Since October
2002 he has been an Associate Professor in the
Dept. of Computer Science and Engineering in the
University Jaume I. His research interests include the optimization
of numerical algorithms for general processors as
well as for specific hardware, and their parallelization
on both message-passing parallel systems (mainly
clusters) and shared-memory multiprocessors. Nowadays
he is involved in several research efforts on HPC
energy-aware systems, cloud computing and HPC
system and development tools.
\end{IEEEbiography}

\vspace{-1cm}
\begin{IEEEbiography}[{\includegraphics[width=1in,height=1.25in,clip,keepaspectratio]{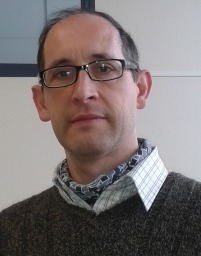}}]{Enrique S. Quintana-Ortí}
received the bachelor’s and PhD degrees in computer sciences from the Universitat Politecnica de Valencia (UPV), Spain, in 1992 and 1996, respectively. After spending the last 23 years as Assistant, Associate and Full Professor at Universidad Jaume I, also in Spain, he recently came back to UPV as Full Professor in Computer Architecture. He has published 350+ papers in international conferences and journals, and has contributed to software libraries such as PLiC/SLICOT, MAGMA, FLARE, BLIS, and libflame for control theory and parallel linear algebra. He has been awarded by NVIDIA and the USA NASA for this development on high performance computing and fault tolerance. His current research interests include parallel programming, linear algebra, energy consumption, transprecision computing and bioinformatics, and advanced architectures and hardware accelerators, and deep learning technologies.
\end{IEEEbiography}

\vspace{-1cm}
\begin{IEEEbiography}[{\includegraphics[width=1in,height=1.25in,clip,keepaspectratio]{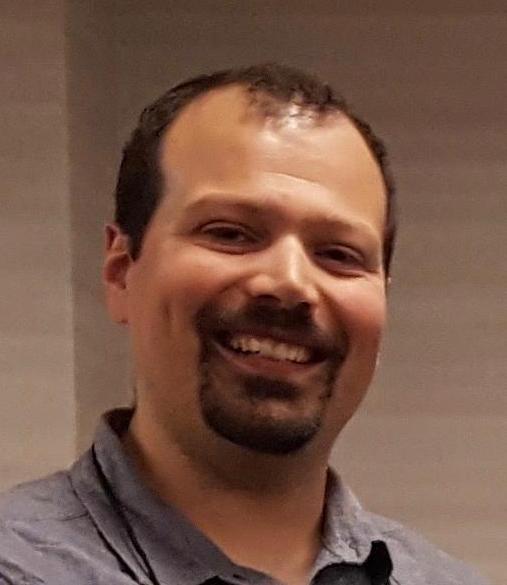}}]{Antonio J. Pe\~na}
 holds a BS+MS degree in Computer Engineering (2006), and
MS and PhD degrees in Advanced Computer Systems (2010, 2013), from
Universitat Jaume I, Spain. He is currently a Sr.\
Researcher at Barcelona Supercomputing Center (BSC), Computer Sciences
Dept. Antonio works within the Programming Models group
where he is Activity Leader for ``Accelerators and Communications for
HPC''. Dr.\ Pe\~na is also the Manager of the BSC/UPC NVIDIA GPU Center of
Excellence.
\end{IEEEbiography}

\newpage

\appendices
\section{API definitions}\label{app:api}

\subsection{DMRlib}
Currently, DMRlib is composed of the following interfaces:
\begin{itemize}
    \item Macro
    \begin{itemize}
        \item DMR\_RECONFIG
    \end{itemize}
    \item Core functions
    \begin{itemize}
        \item DMR\_Set\_paramenters
        \item DMR\_Set\_sched\_period
        \item DMR\_Set\_sched\_itertions
    \end{itemize}
    \item Data transfer functions
    \begin{itemize}
        \item DMR\_Send\_expand\_default
        \item DMR\_Recv\_expand\_default
        \item DMR\_Send\_shrink\_default
        \item DMR\_Recv\_shrink\_default
        \item DMR\_Send\_expand\_blockcyclic
        \item DMR\_Recv\_expand\_blockcyclic
        \item DMR\_Send\_shrin\_blockcyclic
        \item DMR\_Recv\_shrink\_blockcyclic
    \end{itemize}
        \item Variables
    \begin{itemize}
        \item DMR\_INTERCOMM
    \end{itemize}
\end{itemize}

\subsection{DMR API}
The DMR API is defined by:
\begin{itemize}
    \item Functions
    \begin{itemize}
        \item DMR\_Reconfiguration
        \item DMR\_Detach
    \end{itemize}
    \item Compiler directive:
        \begin{itemize}
        \item \#pragma omp task onto
        \end{itemize}
\end{itemize}

\vfill\eject
\section{Energy Consumption}\label{app:energy}
\begin{figure}[h]
  \centering
  \includegraphics[clip,width=1.0\columnwidth,trim={0cm 0cm 0cm 0cm}] {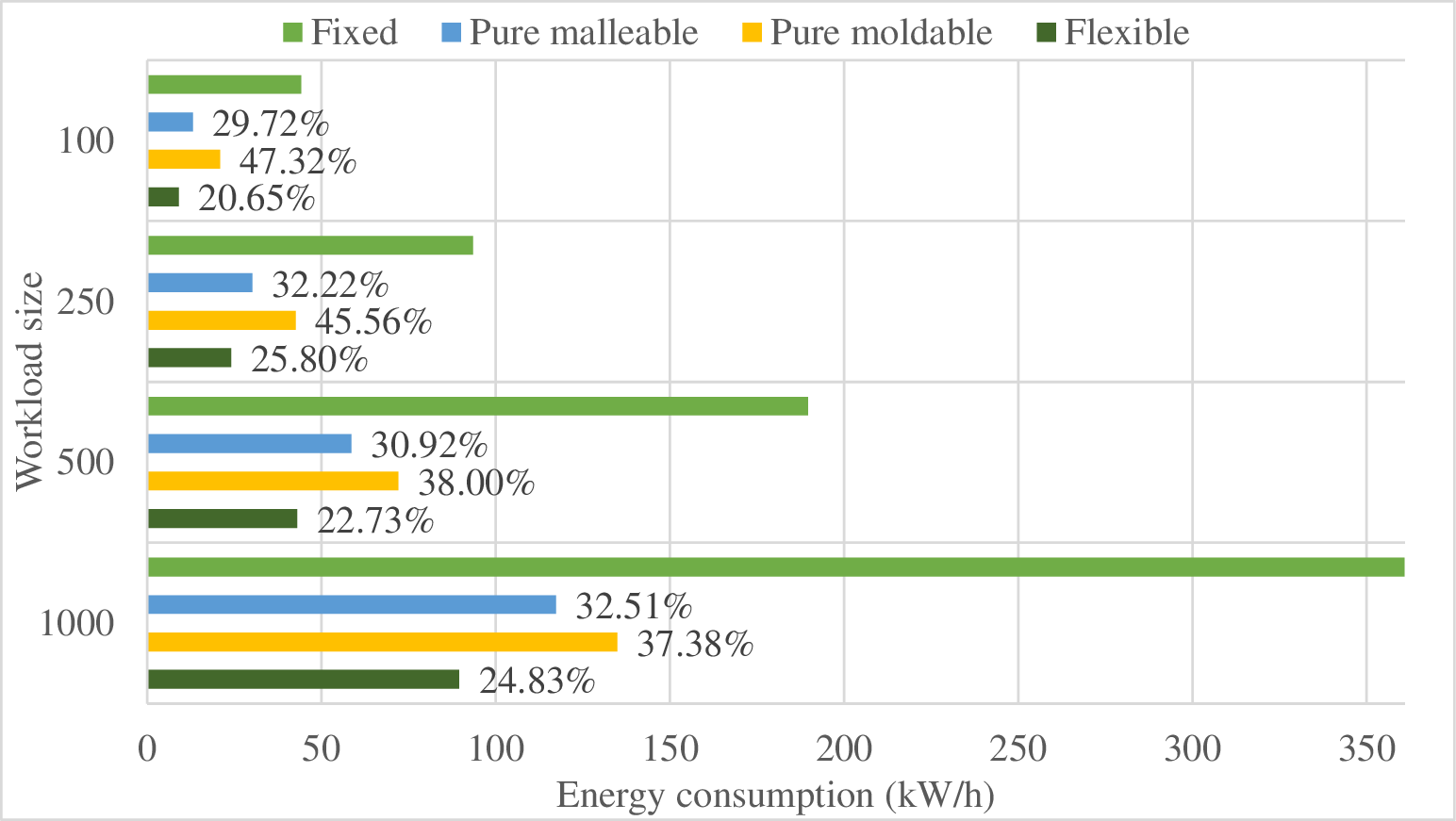}
  \caption{Energy needed to complete a workload compared to the fixed mode.}\label{fig:energy}
\end{figure}

The reduction in time and the level of resource utilization derived from the adoption of DMRlib yield a strong positive impact on energy consumption.
Figure~\ref{fig:energy} depicts the energy usage of each configuration grouped by workload size.
For clarity, we have skipped the 2,000-job workload because it does not provide additional information.
In this figure, the top bar of each size is the reference consumption representing the KW/h of a non-malleable workload with a rigid submission of jobs.
The remaining bars in the group display a label with the relative consumption with respect to the reference value.

From these results we observe that DMRlib can reduce the energy consumption by up to 70\% (second bar of each group).
Furthermore, by changing the submission and using the moldable method, the consumption already decreases to 40--50\% (third bar) of the original level.
Last, the largest reduction is given by combining malleable jobs with moldable submissions, showing that workloads can be then processed with a reduction in the consumption of almost 80\%.

The energy was estimated using the consumption information provided by the technical support of Marenostrum IV: idle nodes consume 100~Wh, while loaded nodes consume 340~Wh.
Energy is then simply calculated by taking into account the time each node is idle/loaded.

\end{document}